\documentclass[article,12pt,a4]{JHEP3}

\usepackage{epsfig}
\usepackage{longtable}
\usepackage{amsmath}
\usepackage{amssymb,amsfonts}

\renewcommand{\theequation}{\arabic{section}.\arabic{equation}}
\def\II{\relax{\rm I\kern-.18em I}}
\def\be{\begin{equation}}
\def\ee{\end{equation}}
\def\bs{\begin{subequations}}
\def\es{\end{subequations}}
\def\bc{\begin{center}}
\def\ec{\end{center}}

\newcommand{\een}{\end{subequations}}
\newcommand{\ben}{\begin{subequations}}
\def\beq{\begin{equation}}
\def\eeq{\end{equation}}

\def\IC{\mathbb{C}}

\def\IR{\mathbb{R}}
\def\IZ{\mathbb{Z}}

\def\II{{\cal I}}

\def\LL{{\cal L}}
\def\MM{{\cal M}}
\def\NN{{\cal N}}
\def\OO{{\cal O}}

\def\SS{{\cal S}}

\def\VV{{\cal V}}
\def\WW{{\cal W}}

\def\p{\partial}

\newcommand\fverb{\setbox\pippobox=\hbox\bgroup\verb}
\newcommand\fverbdo{\egroup\medskip\noindent%
                        \fbox{\unhbox\pippobox}\ }
\newcommand\fverbit{\egroup\item[\fbox{\unhbox\pippobox}]}
\newbox\pippobox

\def\beq{\begin{equation}}
\def\eeq{\end{equation}}
\newcommand{\bea}{\begin{eqnarray}}
\newcommand{\eea}{\end{eqnarray}}

\def\4R{{{}^{(4)}R}}

\def\K5{{\kappa}}
\def\K52{{\kappa^2}}

\title{Josephson Junctions and AdS/CFT Networks}
\author{\href{http://hep.physics.uoc.gr/~kiritsis/}{Elias Kiritsis}$^{1,2}$ 
and Vasilis Niarchos$^{1 \dagger}$
\\ \hspace*{\fill}
\\$^{1}$\href{http://hep.physics.uoc.gr/}{Crete Center for Theoretical Physics},
\\ Department of Physics, University of Crete, 71003, Greece;\\
$^{2}${\href{http://www.apc.univ-paris7.fr/APC_CS/}{Laboratoire APC, Universit\'e Paris-Diderot Paris 7, CNRS UMR 7164, }}\\
{10 rue Alice Domon et L\'eonie Duquet, 75205 Paris Cedex 13, France; }\\
 \hspace*{\fill}
\\$^\dagger${\tt niarchos@physics.uoc.gr}\\ \hspace*{\fill}
}

\preprint{ CCTP-2011-13 \vspace{2.8cm}}

{\bf \abstract{\rm We propose a new holographic model of Josephson junctions (and networks thereof)
based on designer multi-gravity, namely multi-(super)gravity theories on products of distinct
asymptotically AdS spacetimes coupled by mixed boundary conditions. We present a simple model
of a Josephson junction (JJ) that exhibits the well-known current-phase sine relation of
JJs. In one-dimensional chains of holographic superconductors we find that the Cooper-pair
condensates are described by a discretized Schr\"odinger-type equation. Such non-integrable
equations, which have been studied extensively in the past in condensed matter and optics
applications, are known to exhibit complex behavior that includes periodic and quasiperiodic solutions,
chaotic dynamics, soliton and kink solutions. In our setup these solutions translate to
holographic configurations of strongly-coupled superconductors in networks with weak site-to-site
interactions that exhibit interesting patterns of modulated superconductivity. In a continuum limit our
equations  reduce to generalizations of the Gross-Pitaevskii equation. We comment on the many
possible extensions and applications of this new approach.
}


\begin{document}

\section{Introduction}
\label{intro}

There has been recent interest in the potential applicability of AdS/CFT-inspired methods to traditional condensed
matter problems, which are not amenable to a weak coupling quasiparticle description. This has
led to the formulation and study of a large variety of models in classical gravity which aspire
to capture holographically the characteristic features of some condensed matter system.
A notable example is \cite{Gubser:2008px,Hartnoll:2008vx} which provides a simple gravity dual
for an s-wave superconductor. More recent developments in this subject are reviewed in
\cite{Hartnoll:2009sz,Herzog:2009xv,Horowitz:2010gk}. Ultimately one hopes that holographic
techniques will provide a new efficient description of high-$T_c$ superconductors
that goes well beyond the BCS theory.

Many technological applications of superconductors and superconducting devices involve
Josephson junctions (JJs). The basic junction consists of two superconductors separated
by a weak link. The precise type of JJ depends on the specifics of the constituent superconductors
and the nature of the link. The link can be an insulator (SIS junctions), a normal conductor (SNS
junctions) or another superconductor. The coupled superconductors can be of the same or different
type. For example, one can consider $sIs$, $sId$, or $dId$ junctions ($s$ denoting an s-wave
superconductor, $d$ a d-wave superconductor and $I$ an insulator). The properties of these junctions
can be considerably different. For instance, quantum tunneling in conventional SNS and SIS
junctions implies a current $I$ across the link, even in the absence of external voltage, which
depends on the phase difference $\vartheta$ of the condensates of the two superconductors in the
following way \cite{LF}
\beq
\label{introaa}
I=I_{\rm max} \sin\vartheta
~.
\eeq
This simple sinusoidal relation can be substantially different in other (more general) types of
junctions (see for example \cite{Haberkorn,yip2,yip3,yip1,tanaka}).
 
Another reason to be interested in JJs is the nature of high-$T_c$ superconductivity itself.
Many high-$T_c$ superconductors enjoy a layered structure \cite{kivelson} that can be
viewed as a natural stack of atomic scale intrinsic JJs \cite{kleiner} with interlayer spacing
of about 15.5 $\mathring{\rm A}$. In fact, such high quality SIS-type intrinsic JJs can be
fabricated \cite{krasnov1,krasnov2} and pose as attractive candidates of cryoelectronics.

Therefore, assuming holography can provide a new window to high-$T_c$ superconductor physics,
there is an obvious interest to construct the holographic dual of Josephson junctions and
more generally the dual of Josephson junction networks (JJNs).

A first step towards the construction of a holographic SNS junction has been taken recently in
Ref.\ \cite{Horowitz:2011dz} (see also \cite{Keranen:2009ss,Aperis:2010cd,Flauger:2010tv,
Keranen:2010sx,Wang:2011rv,Siani:2011uj} for related setups). 
In this approach one is looking for solutions of the equations
of motion of the standard holographic superconductor setups that are inhomogeneous across
one of the field theory directions (the direction along which the SNS stack is arranged). In the
present paper we will propose a distinctly different way to construct a holographic Josephson
junction (one that is not necessarily restricted to SNS types).

The basic idea is to view each of the superconductors that compose the junction as a separate
supergravity (or superstring) theory on its own asymptotically AdS spacetime and to
model the weak link between them as a mixed boundary condition that relates the boundary
conditions of the condensing symmetry-breaking field on one spacetime to the boundary
conditions of the condensing symmetry-breaking field on the other spacetime. This may look
like a contrived operation on the gravitational side but it is a rather natural one from the perspective
of a dual large-$N$ quantum field theory. On the field theory side this operation amounts to a
multi-trace deformation that involves products of single-trace operators from both theories. This
multi-trace interaction is the only term that mediates interactions between the two theories.
The single-trace operators that appear in these interactions are charged under the broken
$U(1)$ symmetries and thus mimic naturally the charge quantum tunneling effects that are
present in a JJ.

Using multi-trace interactions to model the tunneling effects is natural in a regime where
the mass scales of the modes that mediate the interactions between the two superconductor field
theories in the full system are large compared to the typical energy scales that we consider. In that case
one can integrate out these higher mass modes to obtain an effective theory at low energies.
The separate gauge invariance in each of the two boundary theories implies that the effective
interactions between them can only be of the multi-trace type. Abstracting from this picture the
main features of the bulk description we will proceed to employ them freely in more generic situations
where explicit knowledge about the boundary description is very limited or altogether absent.

The precise ingredients of our construction are presented in section \ref{network}. A simple
characteristic example of a holographic JJ at zero temperature is discussed in section \ref{dimer}.
We show that the standard sine expression for the Josephson current \eqref{introaa} is naturally
reproduced in this model in a few lines and determine $I_{\rm max}$ explicitly in terms of the
parameters of the system. We briefly comment on the extensions and modifications that can
alter this standard current-phase relation. We also discuss how this system differs from a typical 
Josephson junction and what kind of extensions can be used to describe more typical Josephson 
junctions.

Another appealing feature of the above approach is its versatility in describing very general
configurations of Josephson junction networks. JJNs is a much studied subject with diverse
applications. The great wealth of possibilities that they pose and the great reliability of the fabrication
technologies developed for their construction makes them a prototype of complex physical
systems that exhibits a variety of interesting physical behaviors. Among their many applications:
they are used widely as microwave sources (see $e.g.$ \cite{pagano}),
they provide controllable settings to investigate properties of granular or high-$T_c$ superconductors
\cite{simanek}, they are frequently used as model analogs of physical systems with complex dynamics.
For instance, they have been used to model biologically realistic neurons \cite{crotty}.

As a first simple application of our proposal in this direction we consider in section \ref{arrays}
a holographic Josephson junction array that can also be viewed as a special case of the honey-comb
network of Ref.\ \cite{sodano}. Using the gravitational description we find that the Cooper-pair
condensates are described by a discretized Schr\"odinger-type equation, which has been studied
extensively in the past (for a special value of one of our parameters) in radically different
condensed matter and optics applications \cite{HT}. Using well known facts about this equation we
show that the system in question exhibits complex behavior that includes periodic and quasiperiodic
solutions, chaotic dynamics, solitons and kink solutions. In our setup these solutions
translate to one-dimensional configurations of holographic superconductor layers that exhibit
interesting patterns of modulated superconductivity. In a continuum limit, the discretized Schr\"odinger
equation becomes naturally a generalization of the Gross-Pitaevskii (GP) equation, a well-known
long-wavelength description of superfluids. In this limit we recover some of the previously discrete
solutions analytically.

We conclude in section \ref{discussion} with an outline of further possible extensions and applications.

\section{Networks of large-$N$ QFTs and designer multigravity}
\label{network}

\subsection{Networks of large-$N$ QFTs}

Consider a set (network) of $k$ $d$-dimensional QFTs with a large-$N$ limit ---for example,
$k$ potentially different large-$N$ conformal field theories (CFTs). We will label each
CFT by an index $i$ ($i=1,\ldots, k$). Equivalently, $i$ is an index that labels a site (vertex) in our
network. The links of the network are provided by interactions coupling the CFT$_i$'s with each
other. The only kind of coupling respecting the individual gauge structure of each site is one
mediated by multi-trace operators, so this is the only kind that we will consider here. For example,
if $\OO_i$ is a single-trace operator in the CFT$_i$, then a double-trace link between the CFT$_i$ and
the CFT$_j$ corresponds to a Lagrangian interaction of the form
\beq
\label{aaa}
\delta \LL=h \, \OO_i \OO_j
~.
\eeq
Assuming that the large-$N$ scaling of the single-trace operators $\OO_i$ goes like $\OO(N)$,
the double-trace interaction \eqref{aaa} respects the large-$N$ expansion when $h$ is taken
to scale as a constant, namely $h\sim \OO(N^0)$. Similarly, an $\ell$-trace coupling that
involves the product $\prod_{s=1}^\ell \OO_{i_s}$ should have a coefficient that scales as
$\OO(N^{2-\ell})$.

\FIGURE[t]{
\vspace{.4cm}
\centerline{\includegraphics[width=10.3cm,height=7.4cm]{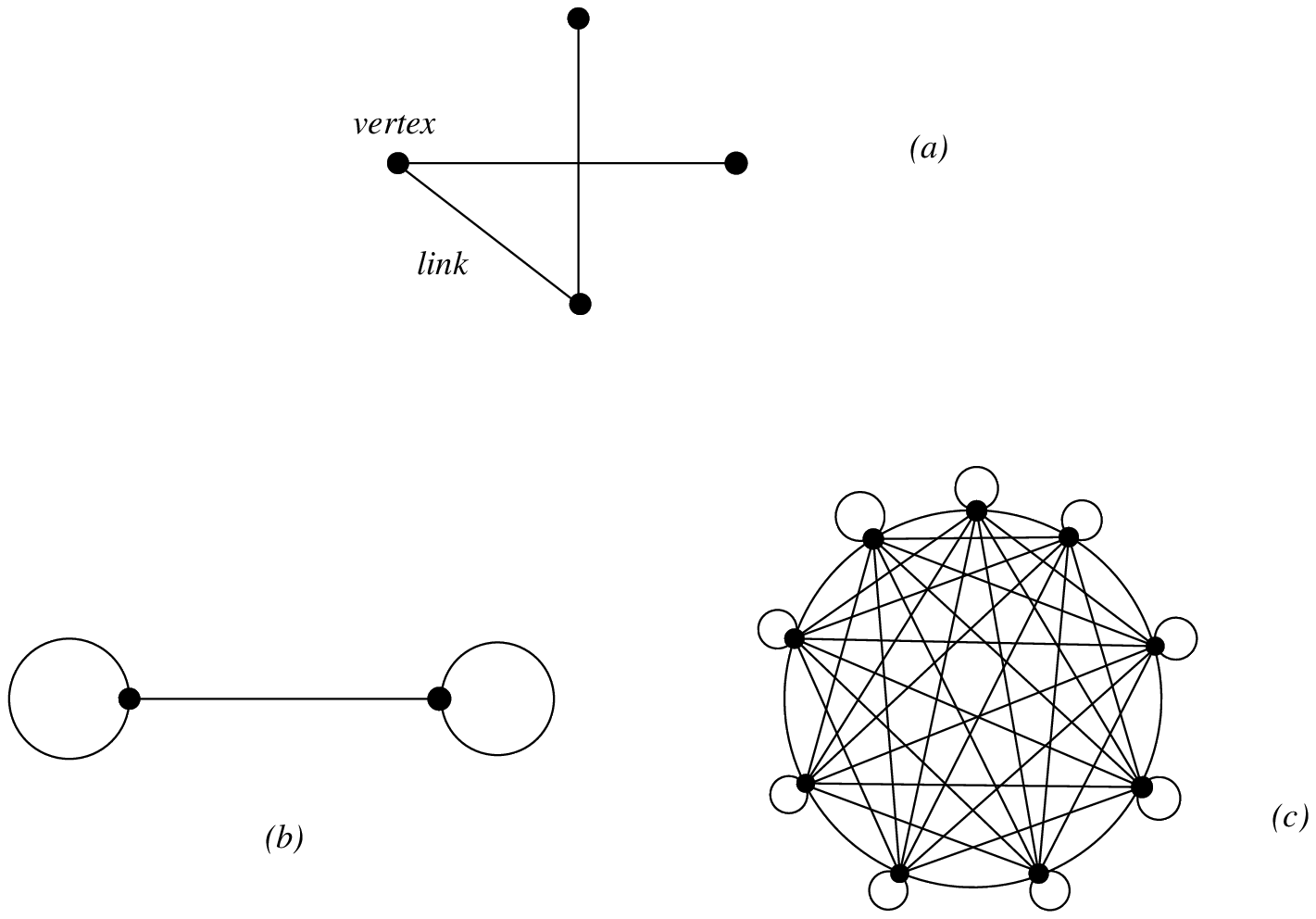}}
\caption{\small \it Examples of simple networks. In our context, these graphs will represent
large-$N$ CFTs coupled by double-trace interactions.}
\label{networks1}
}

It will be useful to set up a convenient notation to denote graphically networks constructed in this
way. For CFTs coupled by double-trace interactions, like that in eq.\ \eqref{aaa}, we will denote the
corresponding link by a single line. Such a link is undirected and joins two different CFTs,
or circles back to the same CFT. The latter denotes that the corresponding CFT has itself a
double-trace interaction turned on. Hence, Fig.\ \ref{networks1}(b) exhibits a simple network of two
CFTs, call them CFT$_1$ and CFT$_2$, with the Lagrangian interaction
\beq
\label{aab}
\delta \LL=h_{11}\OO_1^2+h_{12} \OO_1\OO_2+h_{22} \OO_2^2
~.
\eeq
Similarly, Fig.\ \ref{networks1}(c) exhibits nine CFTs coupled pairwise by double-trace interactions
in all possible ways. This example, analyzed in \cite{Kiritsis:2008at},  plays a role in quenched disorder calculations using the replica
trick \cite{Fujita:2008rs}.

Networks formulated in this way can have different types of links. We
may use a double-line notation to denote the more general possibility of links mediated
by multi-trace interactions. Figs.\ \ref{networks2}(a) and \ref{networks2}(b) exhibit a triple-trace
and four-trace coupling respectively. In network literature the corresponding graphs are
sometimes called hypergraphs. The use of such more general couplings opens up many
interesting possibilities. Most of our discussion in this paper will be focused, however, on
networks with double-trace couplings only.

\FIGURE[t]{
\vspace{.4cm}
\centerline{\includegraphics[width=9.8cm,height=8.2cm]{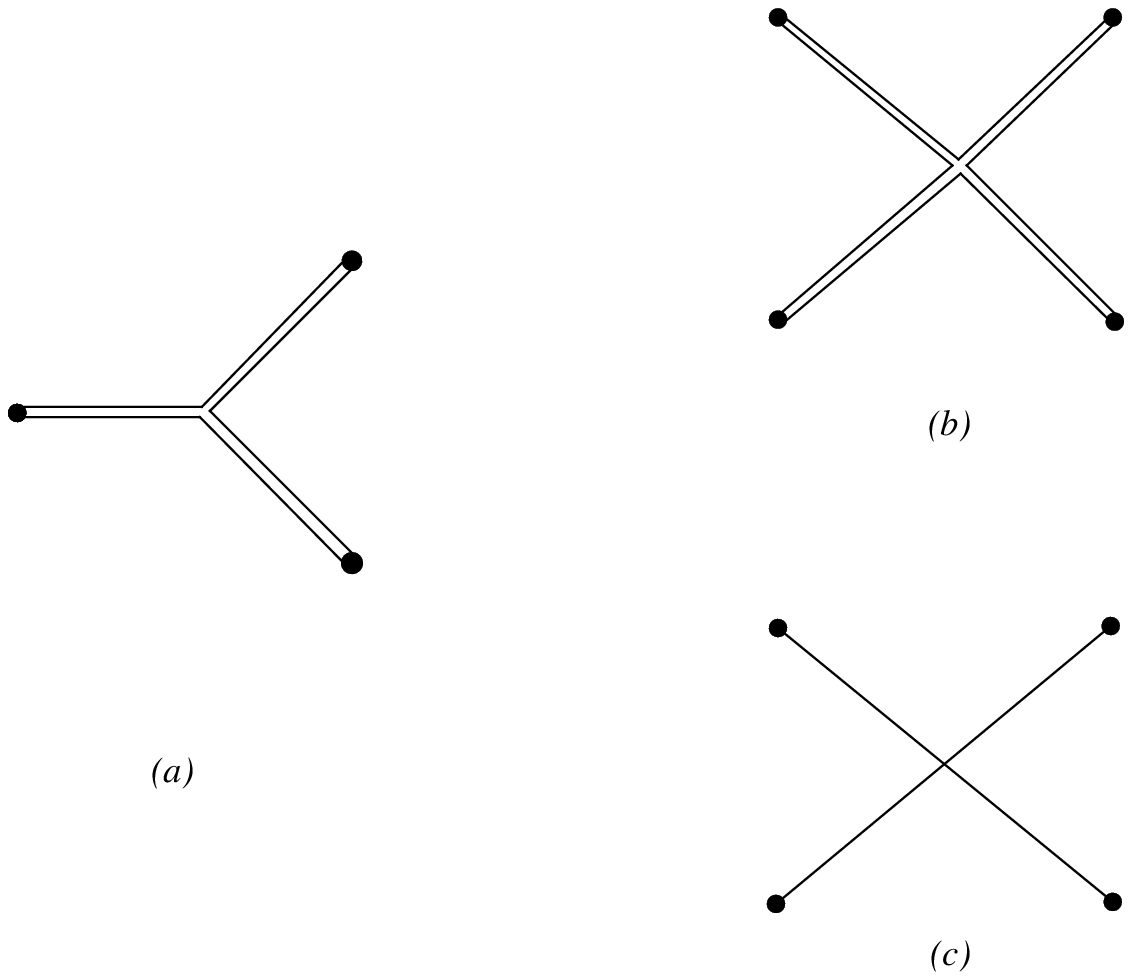}}
\caption{\small \it Figures (a) and (b) exhibit a triple-trace and four-trace link respectively.
In order to make the notation more transparent we have denoted a multi-trace link using a
double-line notation reserving the single-line notation for the simpler double-trace link.
Accordingly, Figure (c) exhibits four CFTs linked pairwise with two double-trace interactions
whereas Figure (b) exhibits a four-trace link.}
\label{networks2}
}

Being comprised by sites that correspond to interacting QFTs the above networks have in general a rich
and complicated internal structure. This structure can evolve in time, vary in space, or
vary from site to site. Moreover, under renormalization group flow the network graphs may
change with the appearance of new links or even new sites. Indeed, it is well-known that
multi-trace couplings are naturally generated under renormalization (see for instance
\cite{Kiritsis:2008at}). Hence, renormalization effects in field theory can affect
the quantitative features of the links. They can also change the number of sites in the
following way. Many supersymmetric QFTs, like the $\NN=4$ super-Yang-Mills theory,
have a Coulomb branch. At generic points of this branch the original gauge group is Higgsed into
product gauge groups and in the far infrared one is left with a product of QFTs giving rise to a
network with more sites than those of the UV theory. This is an example of an RG flow with a
different number of sites in the UV and IR. In what follows, when we draw a graph
representing a network we will implicitly assume that this description refers to the bare UV
Lagrangian of the corresponding theory.

\subsection{Networks of asymptotically AdS spacetimes}

There are situations where a large-$N$ QFT (typically at strong coupling) has a dual description
in terms of a supergravity theory on an asymptotically AdS background. Accordingly, a network
of such QFTs has a dual description as a multi-gravity network where each site is some
supergravity theory on an asymptotically AdS background and each link is a mixed boundary
condition for supergravity fields residing on different space-times \cite{Kiritsis:2006hy,Aharony:2006hz,
Kiritsis:2008at} (see \cite{Kiritsis:2008xj} for a stringy setup that involves multi-string theory networks
and \cite{Niarchos:2009qb} for a brief review of the main idea and its implications for massive
multi-gravity). Let us recall how such a multi-gravity network comes about in the AdS/CFT
correspondence. For clarity, we will restrict to the case where the boundary QFTs are conformal.

Before the addition of multi-trace links, the CFT$_i$s are independent field theories that do not talk
to each other and the dual gravitational theory is a direct product of supergravity theories on
product AdS space-times of the form $\prod_{i=1}^k [AdS_i\otimes \MM_i]$, where $\MM_i$ is
the internal manifold of the spacetime of the dual of CFT$_i$.

Assume that the single-trace operators $\OO_i$ stated above are scalar
operators\footnote{It is not necessary to restrict ourselves to scalar operators. We will make
this assumption here for reasons of simplicity and concreteness. In fact, some of the applications
of this framework that we will propose later also involve vector operators.} with scaling dimension
$\Delta_i$. The AdS/CFT correspondence maps each $\OO_i$ to a dual scalar field $\varphi_i$
with AdS asymptotics
\beq
\label{aba}
\varphi_i \simeq \frac{\alpha_i}{r_i^{\Delta_i}}+\ldots+\frac{\beta_i}{r_i^{d-\Delta_i}}+\ldots
~.
\eeq
$r_i$ is the radial distance in the $i$-th AdS space that corresponds to CFT$_i$.
We use conventions where the $i$-th AdS boundary lies at $r_i \to \infty$.
In later applications, we will assume $\Delta_i<\frac{d}{2}$, in which case the $\alpha_i$'s
should be interpreted as the vacuum expectation values (VEVs) of the dual operators
and the $\beta_i$'s as the sources.

In this framework, and to leading order in the $1/N$ expansion, it is well known that a
multi-trace coupling in the boundary CFT
\beq
\label{abb}
\delta \LL=\WW(\OO_1,\ldots, \OO_k)
\eeq
imposes mixed boundary conditions to the asymptotic coefficients $\alpha_i,\beta_i$ of the form
\beq
\label{abc}
\beta_i=\p_i \WW(\alpha_1,\ldots,\alpha_k)
~.
\eeq
Combined with requirements of regularity these boundary conditions fix completely the profile
of the bulk solution. The inter-theory coupling induced by the relations \eqref{abc}
leads to a network of supergravity theories that can exhibit interesting collective phenomena.
We will explore these phenomena in section \ref{arrays}.

\subsection{Designer multigravity}

Let us recall how multi-trace deformations affect the profile of the bulk asymptotically AdS solution
in a single theory; a subject that usually goes under the title of designer gravity \cite{Hertog:2004ns}.
In this paper we want to consider the natural extension of this framework to multi-AdS spaces, which
we will suggestively call `designer multigravity'.

As one of the simplest illustrations of the idea consider the `network'
\beq
\label{condaa}
\begin{array}{c}
\includegraphics[width=2cm,height=2cm]{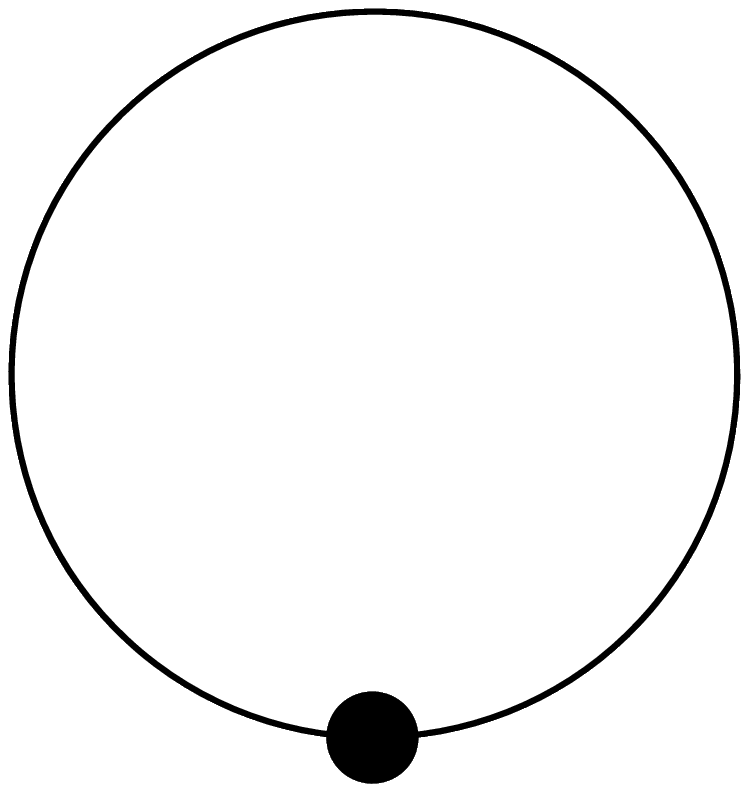}
\end{array}
\eeq
which contains a single vertex with a self-adjoining link. The graph \eqref{condaa} exhibits a
double-trace link, but we can equally well consider any multi-trace link. This theory has been the main
focus of most previous investigations of multi-trace interactions in the AdS/CFT correspondence and
designer gravity. For self-completeness and in order to set the notation, we briefly review some
of the most pertinent properties of this system.

At the single site of this network resides a $d$-dimensional large-$N$ CFT with a supergravity
dual. Assume that the CFT has a single-trace complex scalar operator $\OO$ with scaling dimension
$\Delta$. This
operator is dual to a charged bulk scalar field $\varphi$. To further simplify the discussion we will
also assume that we can consistently reduce the dynamics of the dual bulk supergravity to a
$(d+1)$-dimensional Einstein-Abelian Higgs model of the form
\beq
\label{condab}
S_{bulk}=\int d^{d+1}x\, \sqrt{-g} \, \left[ R-\frac{1}{4} G(|\varphi|) F^2
-(\nabla |\varphi|)^2-J(|\varphi|) \left( \nabla \theta-q A \right)^2 -  V(|\varphi|) \right]
\eeq
where $\theta$ is the phase of $\varphi$, namely $\varphi=|\varphi| e^{i\theta}$, and
$G, J, V$ are model-dependent functions of $|\varphi|$. $A$ is an Abelian gauge field in the bulk
and $F$ its field strength. $q$ is the $U(1)$ charge of the field $\varphi$.

We are interested in asymptotically AdS$_{d+1}$ solutions of this system. In units where the
AdS radius is set to one, the potential $V(\varphi)$ has the small-$\varphi$ expansion
\beq
\label{condac}
V(|\varphi|)=-d(d-1)+m^2 |\varphi|^2+\OO(|\varphi|^4)+\ldots
~.
\eeq
When the mass $m$ lies within the range
\beq
\label{condad}
m^2_{BF}<m^2<m^2_{BF}+1~, ~~ m^2_{BF}=-\frac{d^2}{4}
\eeq
the dual operator $\OO$ can have two possible scaling dimensions
\beq
\label{condae}
\Delta_\pm=\frac{d}{2}\pm \sqrt{\frac{d^2}{4}+m^2}
~.
\eeq
We will assume that our theory lies in the window \eqref{condad} and will pick $\OO$ to have
the smaller scaling dimension $\Delta\equiv \Delta_-$ that satisfies the inequality
$\frac{d-2}{2}<\Delta_-<\frac{d}{2}$ (the lower bound in this inequality is the standard unitarity
bound in field theory).

Near the asymptotic boundary, $r\to \infty$, the metric is that of AdS$_{d+1}$
\beq
\label{condafa}
ds^2\simeq r^2 dx_\mu dx^\mu+\frac{dr^2}{r^2}
\eeq
and the scalar field
$\varphi$ exhibits two independent branches
\beq
\label{condaf}
\varphi\simeq \frac{\alpha}{r^\Delta}+\ldots +\frac{\beta}{r^{d-\Delta}}+\ldots
~.
\eeq
The boundary condition
\beq
\label{condag}
\beta=\frac{d \WW}{d\alpha}
~,
\eeq
for a generic smooth function $\WW(\alpha)$ corresponds at the boundary CFT to the multi-trace
deformation \cite{Witten:2001ua}
\beq
\label{condai}
\delta \LL= \WW(\OO)
~.
\eeq

When we solve the bulk equations of motion we are looking for solutions that respect the
boundary conditions \eqref{condag}. In addition, we require that these solutions are
everywhere regular. It turns out that regularity imposes an extra constraint on the
asymptotic coefficients $\alpha$, $\beta$. We will denote this
additional relation as
\beq
\label{condaj}
\beta=-\frac{d\widetilde \WW}{d\alpha}
\eeq
where $\widetilde \WW$ is a function determined by the specific dynamics of the theory, $e.g.$
the details of the scalar potential $V(|\varphi|)$ in the bulk action \eqref{condab}.

Combining the boundary condition \eqref{condag} and the regularity condition \eqref{condaj}
we find that $\alpha$ and $\beta$ are completely fixed and determined as a solution to the equation
\beq
\label{condak}
\frac{d\VV}{d \alpha}=0~, ~~ \VV(\alpha)=\WW(\alpha)+\widetilde \WW(\alpha)
\eeq
which therefore can be viewed as an extremum of the function $\VV$.

For example, for boost invariant, planar configurations with vanishing gauge field
\beq
\label{condal}
ds^2=r^2\left(-dt^2+dx_i dx^i\right)+\frac{dr^2}{g(r)}~, ~~ \varphi=\varphi(r)~, ~~ A=0
\eeq
one finds a solution with an acceptable naked singularity at $r=0$ that has
\beq
\label{condam}
\widetilde \WW(\alpha)=s\frac{\Delta}{d}|\alpha|^{d/\Delta}
~.
\eeq
The existence of this solution and the precise value of the parameter $s$ depends on the details
of the bulk potential $V(\varphi)$ (see \cite{Faulkner:2010fh} for additional information).
For most potentials, $s$ turns out to be a positive number. The energy density of this solution is
\beq
\label{condan}
\frac{E}{Vol}=(d-2\Delta)\VV
~.
\eeq

One can also consider analogous solutions with spherical topology. It is possible to show
that the energy \eqref{condan}, \eqref{condam} provides a lower energy bound for all of these
solutions \cite{Faulkner:2010fh}. Moreover, in the case of vanishing gauge field, one can argue
\cite{Faulkner:2010fh} that
\begin{itemize}
\item the theory with boundary conditions $\beta=\frac{d\WW}{d\alpha}$ has a stable ground
state provided the function $\VV$ has a global minimum $\VV_{min}$, and
\item that the minimum energy solution is the spherical soliton associated with $\VV_{min}$.
\end{itemize}

Note that because of the presence of $\widetilde \WW$, it is possible to have a stable ground state
even for functions $\WW$ that have no minimum. In general, this minimum
involves a condensate of the charged scalar that higgses the corresponding $U(1)$.

What we have discussed so far applies to the case of zero temperature. By studying hairy black holes
in the bulk it is possible to generalize the discussion to non-zero temperature \cite{Hertog:2005hu}. It is
also possible to consider the case of non-vanishing charge density \cite{Faulkner:2010gj}.

It is not hard to generalize this discussion to the case of multiple CFTs coupled together by
a multi-trace interaction of the general form $\WW(\OO_1,\ldots, \OO_k)$. In this case
(and to leading order in the $1/N$ expansion), the bulk
equations of motion are the same as before in each (super)gravity member of the product,
but the boundary conditions change. For example, the bulk boost-invariant planar solutions in the
$i$-th spacetime still retain the form \eqref{condal} and each $\beta_i$ is still given by an equation
of the form
\beq
\label{condaoa}
\beta_i = -\frac{d \widetilde W_i(\alpha_i)}{d\alpha_i}
\eeq
with
\beq
\label{condao}
\widetilde W_i(\alpha_i)= s_i \frac{\Delta_i}{d} |\alpha_i|^{d/\Delta_i}
~.
\eeq
The new ingredient, responsible for the coupling between different AdS theories,
lies in the mixed boundary conditions \eqref{abc}. The analog of equation \eqref{condak}
that determines the VEVs $\alpha_i$ is
\beq
\label{condap}
\frac{\p \VV(\alpha_1,\alpha_2,\ldots, \alpha_k)}{\p \alpha_i}=0
\eeq
with
\beq
\label{condaq}
\VV=\WW (\alpha_1,\ldots,\alpha_k)+\sum_{i=1}^k \widetilde \WW_i(\alpha_i)
~.
\eeq
In general, these equations lead to a non-linear discrete system which can exhibit
intricate behavior. This behavior includes solutions that can be periodic, quasi-periodic,
chaotic or even soliton-like with energy pinned around a central site. Examples of
each of these behaviors will be discussed in section \ref{arrays}.

\subsection{A holographic superconductor with vanishing charge density}
\label{example}

A novel type of holographic superconductor with symmetry breaking induced by double-trace
deformations was recently proposed in \cite{Faulkner:2010gj}. Since this setup will provide the
basic building block of the discussion to come, it will be beneficial to recall some of its main properties.

Returning to the single-site example of \eqref{condaa} consider the case of a double-trace
deformation
\beq
\label{condba}
\WW=g |\OO|^2
\eeq
implemented with the use of a single-trace operator $\OO$ with scaling dimension $\Delta$.
The function $\VV$ in \eqref{condak} becomes in this case
(for boost-invariant planar solutions)
\beq
\label{condbb}
\VV(\alpha)=g |\alpha|^2+\frac{s}{\delta} |\alpha|^\delta~, ~~ \delta\equiv \frac{d}{\Delta}
~.
\eeq
The extrema of this function obey the algebraic equation
\beq
\label{condbc}
\alpha \left( g+\frac{s}{2} |\alpha|^{\delta-2}\right)=0
~.
\eeq

Assuming $s>0$, there is one or two possible solutions to this equation depending on the sign of
the constant $g$. For $g>0$, the only solution is $\alpha=0$, which is a minimum and does
not exhibit any condensate of the field $\varphi$. For $g<0$, there are two possible
solutions
\beq
\label{condbd}
\alpha_0=0~,  ~~~\alpha_\theta=\left( -\frac{2g}{s} \right)^{\frac{1}{\delta-2}}e^{i\theta}
~.
\eeq
The first one, $\alpha_0$, is a local maximum of $\VV$ and therefore an unstable vacuum of the
system. The other solutions, $\alpha_\theta$, are degenerate stable minima labeled by an angular
variable $\theta$. The non-vanishing condensate of the charged scalar field $\varphi$ in this
case higgses the $U(1)$ gauge symmetry and leads to a new type of holographic superconductor.

\section{A holographic model of Josephson junctions}
\label{dimer}

\subsection{The setup}
\label{setdimer}

A Josephson junction consists of two superconductors separated by a link that mediates
weak interactions between them (several possibilities for the weak link were reviewed in the
introduction). In analogy, consider a setup where two holographic superconductors (each
described by a gravitational theory on an asymptotically AdS space) are interacting weakly via mixed
boundary conditions on the boundary.\footnote{We will soon discuss how similar this system is
to the typical Josephson junction considered in the laboratory and the extensions it suggests.} 
From the gauge theory point of view two initially
separate large-$N$ gauge theories are brought into contact via multi-trace interactions.
As we discussed in the introduction, in certain cases one can think of the multi-trace interactions
as an effective description below the typical mass gap associated to the actual interactions
between the two systems. In a real Josephson junction these would be the weak tunneling
interactions across the material in between the superconductors. 

As a concrete illustration of the idea consider a pair of two identical holographic superconductors
of the type described in the previous subsection \ref{example}. The parameters $g, s, \delta,
d, \Delta$, which will be treated here as phenomenological parameters of the model, are chosen to
be common in the two systems. Each of them has a charged scalar
field $\OO_i$ $(i=1,2)$ and a corresponding dual complex scalar field $\varphi_i$ with
asymptotics \eqref{condaf}. We can think of $\OO_i$ as the `Cooper pair' operator in
the holographic superconductor CFT$_i$. Holographically,
the VEV of these operators are given by the leading branch coefficients $\alpha_i$
in the asymptotic expansion \eqref{condaf}. We assume that both operators have the same
scaling dimension $\Delta<\frac{d}{2}$ and employ the alternative quantization for the dual
scalar fields $\varphi_i$ in the bulk AdS spacetime.

In order to mediate an interaction that exchanges charge between the two theories we couple
them via a double-trace interaction of the form
\beq
\label{dimeraa}
\WW(\OO_1,\OO_2)=h \left( e^{i\vartheta} \OO_1 \OO_2^\dagger + e^{-i\vartheta} \OO_1^\dagger
\OO_2 \right)
~.
\eeq
$h$ is a real number controlling the strength of the interaction and $\vartheta$ is an angular
variable. The latter is an additional tunable parameter of the interaction whose physical meaning
will become clear in a moment.

In the bulk this coupling implies mixed boundary conditions. One can determine the VEVs
of the dual operators $\OO_i$ by solving the scalar-gravity equations of motion. Equivalently,
one can extremize the function $\VV$ (see eqs.\ \eqref{condap}, \eqref{condaq}). In the case at hand
\beq
\label{dimerab}
\VV(\alpha_1,\alpha_2)=\sum_{i=1}^2 \left( g|\alpha_i|^2+\frac{s}{\delta} |\alpha_i|^{\delta} \right)
+ h \left( e^{i\vartheta} \alpha_1 \alpha_2^* + e^{-i\vartheta} \alpha_1^* \alpha_2 \right)
~.
\eeq
Extremizing with respect to $\alpha_1$ and $\alpha_2$ we obtain two algebraic equations
\bea
\label{dimerac}
&& g\alpha_1+ h e^{-i\vartheta} \alpha_2+\frac{s}{2} \alpha_1 |\alpha_1|^{\delta-2}=0~,~~
\nonumber\\
&& g\alpha_2+ h e^{i\vartheta} \alpha_1+\frac{s}{2} \alpha_2 |\alpha_2|^{\delta-2}=0
\eea
which determine the $\alpha_i$'s uniquely in terms of the parameters $g,h,s$ and $\delta$ up to
an overall phase common in $\alpha_1$ and $\alpha_2$. The relative phase between
$\alpha_1$ and $\alpha_2$ is fixed in terms of $\vartheta$
\beq
\label{dimerad}
\vartheta_{12} \equiv \vartheta_2-\vartheta_1=\vartheta \mod~ \pi
~.
\eeq

For instance, when $\delta=4$ \footnote{This is consistent with $\Delta \in
\left(\frac{d}{2}-1,\frac{d}{2}\right)$ iff $\Delta<1$, which is possible only if $d=2,3$ (the main
cases of interest in this paper).} the solutions are
\beq
\label{dimerae}
(1)~~\alpha_1=0~, ~~ (2)_\pm ~~|\alpha_1|^2=\frac{2}{s}\left(\pm h-g \right)~, ~~
(3)_\pm ~~|\alpha_1|^2=-\frac{1}{s}\left( g\pm \sqrt {g^2-4h^2} \right)
~.
\eeq
In all cases
\beq
\label{dimeraf}
\alpha_2=-h^{-1}e^{i\vartheta} \left( g+\frac{s}{2}|\alpha_1|^2 \right) \alpha_1
~.
\eeq
Which solutions survive in cases $(2)$ and $(3)$ depends on the specific range of the
parameters $s,g,h$. Assuming $s>0$ we obtain
\begin{itemize}
\item $|h|<g$: vacuum $(1)$,
\item $-|h|<g<|h|$: vacua $(1)$, $(2)_{{\rm sgn}(h)}$,
\item $-2|h|<g<-|h|$: vacua $(1)$, $(2)_\pm$,
\item $g<-2|h|$: vacua $(1)$, $(2)_\pm$, $(3)_\pm$.
\end{itemize}

Diagonalizing the Hessian of the energy functional $\VV$ (see eq.\ \eqref{condan})
we find that $(1)$ has the eigenvalues
\beq
\label{dimerag}
2(g+h)~, ~~ 2(g-h)
~.
\eeq
Hence, when phases with a non-zero condensate exist $(g<|h|)$ the vacuum $(1)$ of the
normal phase is unstable (one of the eigenvalues of the Hessian is negative). In that case
there is always at least one superconducting vacuum that is stable. Such a phase provides a
holographic model of two weakly interacting superconducting materials at zero temperature.

As a simple illustration, by setting $h=1$ we find that
\begin{itemize}
\item when $-1<g<1$, the only stable vacuum is $(2)_+$,
\item when $-2<g<-1$, there are two stable degenerate vacua $(2)_\pm$, and
\item when $g<-2$, the vacua $(3)_\pm$ are also unstable and the only stable vacua are
again $(2)_\pm$.
\end{itemize}

\subsection{Josephson current}
\label{Jcurrent}

A characteristic feature of conventional Josephson junctions is a transverse supercurrent $I$ which is
related to the condensate phase difference $\vartheta_{12}$ in the following way
\beq
\label{curraa}
I=I_{\rm max}\sin\vartheta_{12}
~.
\eeq

In this section we have considered a system of two sites that represents two infinitely thin
layers of superconducting material at zero charge density coupled through a weak link 
expressed by an interaction of the form \eqref{dimeraa}, which can be suggestively rewritten as
\beq
\label{curraaa}
\WW(\OO_1,\OO_2)=h \left( e^{i\vartheta} \OO_1 \OO_2^\dagger + e^{-i\vartheta} \OO_1^\dagger
\OO_2 \right)=\WW_E+\WW_{J_{ext}}
\eeq
with the definition
\beq
\label{curraab}
\WW_E = h\cos\vartheta \left( \OO_1 \OO_2^\dagger+\OO_1^\dagger \OO_2\right)~, ~~
\WW_{J_{ext}} = i h \sin \vartheta \left( \OO_1 \OO_2^\dagger-\OO_1^\dagger \OO_2\right)
~.
\eeq
$\WW_E$ is an interaction that mediates no interlayer charge transfer. In contrast, 
$\WW_{J_{ext}}$ is an interaction based on the charge-transferring operator 
\beq
\label{curraad}
J_{ext} = i \left( \OO_1 \OO_2^\dagger-\OO_1^\dagger \OO_2\right)
~.
\eeq
The coupling $A$ of this operator in $\WW_{J_{ext}}$, namely
\beq
\label{curraae}
\WW_{J_{ext}}=A J_{ext}~, ~~ A=h \sin\vartheta
\eeq
can be interpreted as a new interlayer background gauge potential component.\footnote{This 
component of the gauge field is not an inherent quantity of the $(d+1)$-dimensional 
theories living in each site of our network. In general, a linear array of sites `deconstructs' an extra
spacetime dimension and quantities, like $A$, associated with this extra direction
arise in the field theory space of the lower-dimensional multi-AdS/CFT network (quiver)
as new interlayer interactions.} In this sense, for non-vanishing $h, \vartheta$ our two-site system 
lies in an external transverse gauge field and $J_{ext}$ is an externally imposed current.

The total current running across the sites of the junction can be determined in the standard way 
from an infinitesimal relative $U(1)$ gauge transformation of the action. For an interaction of the
form \eqref{curraaa} the infinitesimal transformation
\beq
\label{totalcuraa}
\delta\OO_1=i \epsilon \OO_1~, ~~ \delta\OO_2=-i \epsilon \OO_2
\eeq
gives the Lagrangian variation
\beq
\label{totalcurab}
\delta \LL=2i h\epsilon \left( e^{i\vartheta} \OO_1 \OO_2^\dagger-e^{-i\vartheta} \OO_1^\dagger \OO_2
\right) =\epsilon(J_{site~1}-J_{site~2})=2 \epsilon J_{tot}
~.
\eeq
The second equality is the discretized version of the gradient $\p J$ across the interlayer direction.
In the third equality we used charge conservation to set $J_{site~1}=-J_{site~2}=J_{tot}$.

Consequently,
\beq
\label{totalcurac}
J_{tot}=h\left( e^{i\vartheta} \OO_1 \OO_2^\dagger-e^{-i\vartheta} \OO_1^\dagger \OO_2\right) 
~.
\eeq
Hence, in the vacuum governed by the algebraic equations \eqref{dimerac} one finds (at leading order
in the $1/N$ expansion)
\beq
\label{totalcurad}
J_{tot}=h\left( e^{i\vartheta} \alpha_1 \alpha_2^*-e^{-i\vartheta} \alpha_1^* \alpha_2\right)=0
~.
\eeq
This is precisely what one expects. Since our system is kept at zero charge density, in the 
equilibrium state charge cannot flow across the junction between the two sites. 
As a result, irrespective of the initial configuration, once the interaction \eqref{curraaa} is 
turned on the system backreacts and evolves to a new vacuum, which can be conveniently 
determined with the holographic techniques of subsection \ref{setdimer}. In the new vacuum, 
which is characterized by the solutions of the algebraic equations \eqref{dimerac}, the condensate 
phase difference is $\vartheta_{12}=-\vartheta$ and the magnitude of the condensate has adapted 
accordingly and in direct relation to the strength of the interlayer couplings $h,\vartheta$. 

It is interesting to consider the VEV of the externally forced current operator $J_{ext}$ in the 
new vacuum. At leading order in the $1/N$ expansion
\beq
\label{currab}
\langle J_{ext} \rangle = i(\alpha_1^* \alpha_2-\alpha_1 \alpha_2^*)
~.
\eeq
Using eq.\ \eqref{dimeraf} we obtain
\beq
\label{curraba}
\langle J_{ext} \rangle=I_{\rm max} \sin\vartheta = -I_{\rm max} \sin \vartheta_{12}
\eeq
with
\beq
\label{currac}
I_{\rm max}=2h^{-1}|\alpha_1|^2 \left( g+\frac{s}{2} |\alpha_1|^2\right)
~.
\eeq
In the algebraically simple case of $\delta=4$
\beq
\label{currad}
I_{\rm max}=\pm \frac{4}{s}\left( \pm h-g \right)
\eeq
on the vacua $(2)_\pm$ and
\beq
\label{currae}
I_{\rm max}=-\frac{4h}{s}
\eeq
on the vacua $(3)_\pm$ (interestingly in this case $I_{\rm max}$ is independent of $g$).

The vanishing of the total interlayer current $J_{tot}$ implies that the backreaction has created
an equal and opposing `Josephson current' $J_{josephson}$ (due to the condensate phase
difference), which cancels the contribution
of the externally imposed $\langle J_{ext} \rangle$. Specifically,
\beq
\label{curraf}
J_{josephson}=-\langle J_{ext} \rangle=I_{\rm max}\sin \vartheta_{12}
\eeq
in agreement with the expected sine law \eqref{curraa}.

It is clear that the system we have just described is a peculiar Josephson junction unlike the typical
Josephson junction commonly discussed in the literature and engineered in the laboratory. In 
contrast to our system in a typical junction the superconductor components have finite spatial 
thickness in the transverse junction direction, charge can flow in this direction, the condensate 
phase difference is dialed by choice (and not determined dynamically) and a Josephson current arises 
without having to apply a gauge field across the weak link. In section \ref{arrays} we will describe
how such more conventional junctions can be engineered in our framework as simple extensions
of the setup we have described in this subsection.

\subsection{Possible extensions and other current-phase relations}

We briefly comment on a few possible extensions of this framework that may ultimately
lead to different current-phase relations.

A possible alternative is to consider the same superconductor models as above,
but modify the double-trace interactions. For example, one could consider other types
of double-trace interactions, $e.g.$ instead of (or in addition to) \eqref{dimeraa} a
double-trace deformation of the form
\beq
\label{currba}
h(e^{i\omega}\OO_1 \OO_2+e^{-i\omega}\OO_1^\dagger \OO_2^\dagger )
~.
\eeq
This will modify the equations \eqref{dimerac} and the Josephson current \eqref{curraf}
in an obvious fashion.

Another possibility is to consider higher multi-trace deformations. An example of a triple-trace
interaction is
\beq
\label{currbb}
h(e^{i\omega} \OO_1 \OO_2^{2 \dagger}+e^{-i\omega} \OO_1^\dagger \OO_2^2)
~.
\eeq

Finally, it is interesting to consider coupling different types of asymptotically AdS
theories. For example, one can try to couple the holographic $s$-wave superconductor of Ref.\
\cite{Hartnoll:2008vx} to the holographic $p$-wave superconductor of Ref.\ \cite{Gubser:2008wv}.
In the $s$-wave superconductor it is a bulk complex scalar field, dual to a complex scalar operator,
that condenses. In the $p$-wave superconductor there is an $SU(2)$ gauge field in the bulk and the
$U(1) \subset SU(2)$ is broken by the condensation of the remaining two components of the
$SU(2)$ gauge field, which are dual to the corresponding components of an $SU(2)$ current
in the boundary theory. One can consider the possibility of scalar-current double-trace deformations
on the boundary which presumably translate to mixed boundary conditions for the dual scalar
and gauge field components.

\section{Holographic Josephson junction arrays}
\label{arrays}

A lot of theoretical and experimental work has been performed on Josephson junction arrays
(and more generally networks of Josephson junctions). Some of the main motivations
and results in this field were summarized in the introduction. A natural generalization of the
holographic construction of the previous section can be used to model networks of holographic
superconductors.

\FIGURE[t]{
\vspace{.4cm}
\centerline{\includegraphics[width=9.8cm,height=2.2cm]{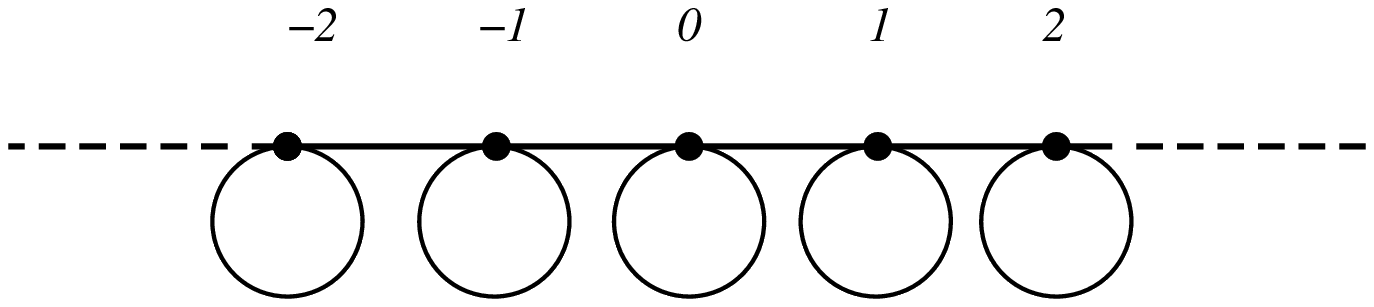}}
\caption{\small \it A chain of holographic superconductors labeled by an index $n\in \IZ$.
Each link denotes an interaction mediated on the field theory side by a double-trace deformation.}
\label{networks5}
}

In what follows we will concentrate on a simple network with the topology of a chain as depicted
in Fig.\ \ref{networks5}. In this network $M$ sites (labeled by an index $n$, each of them representing
a $d$-dimensional holographic superconductor of the type of subsection \ref{example})
are linked by the mixed boundary conditions corresponding to the double-trace interactions
\beq
\label{arraysaa}
\WW(\{ \OO_n \})=g \sum_n \OO_n \OO_n^\dagger+
h\sum_n \left( e^{i\vartheta} \OO_{n} \OO_{n+1}^\dagger+e^{-i\vartheta} \OO_n^\dagger
\OO_{n+1} \right)
~.
\eeq
For simplicity, we are assuming the same constants $g, h,\vartheta,\delta$ for all sites and links.
Accordingly, the potential function $\VV$ that we have to extremize in gravity to obtain the
condensates $\alpha_n$ is
\beq
\label{arraysab}
\VV(\{ \alpha_n \})=\sum_n \left( g |\alpha_n|^2+\frac{s}{\delta}|\alpha_n|^\delta
+ h\left( e^{i\vartheta}\alpha_n \alpha_{n+1}^*+e^{-i\vartheta} \alpha_n^* \alpha_{n+1} \right)
\right)
~.
\eeq
The extrema of this function are sequences of complex numbers obeying recursion relations with
rich features. The solutions can be periodic, quasiperiodic (chaotic) or solitonic, and provide
interesting new examples of spatially modulated, namely lattice site-dependent,
superconductivity.

We will organize the discussion according to the number of boundary conditions.

\subsection{No boundaries}
\label{infinite}

In the case of no boundaries, the extremization of the potential function $\VV$ \eqref{arraysab}
with respect to all the $\alpha_i$'s gives the recursion relations
\beq
\label{arraysba}
g \alpha_n+h \left( e^{i\vartheta} \alpha_{n-1} +e^{-i\vartheta} \alpha_{n+1} \right)
+\frac{s}{2} \alpha_n |\alpha_n|^{\delta-2}=0
~, ~~ n\in \IZ
~.
\eeq
Setting
\beq
\label{arraysbb}
\alpha_n \equiv e^{in \vartheta} \varphi_n  ~, ~~ \tilde g \equiv \frac{g}{h}~, ~~ \tilde s\equiv \frac{s}{h}
\eeq
we can recast \eqref{arraysba} into the form
\beq
\label{arraysbc}
\tilde g \varphi_n+\varphi_{n-1}+\varphi_{n+1}+\frac{\tilde s}{2} \varphi_n |\varphi_n|^{\delta-2}=0
~.
\eeq

The generic solution of these equations is parameterized by two complex numbers (these
could be for instance the values of $\varphi_0$, $\varphi_1$ at the vertices 0 and 1).
There are various ways to analyze this system of equations. In fact, such equations have
appeared in the past in a variety of applications and discussions of discrete dynamical systems.
A notable example, that arises for the special case of $\delta=4$, is that of the discrete non-linear
Schr\"odinger (DNLS) equation
\beq
\label{arraysbd}
-i \frac{d}{dt}\psi_n+\psi_{n-1}+\psi_{n+1}+\frac{s}{2} \psi_n |\psi_n|^2=0
~.
\eeq
This equation is a discretized version of the Schr\"odinger equation with a non-linear
quartic potential. Setting
\beq
\label{arraysbe}
\psi_n \equiv \varphi_n e^{i E t}
\eeq
we recover our set of equations \eqref{arraysbc} with $E=\tilde g$ and $\delta=4$.

The DNLS equation has a long history (for an extensive review and references we refer the reader
to \cite{HT}). In solid state physics it appeared first in the context of the Holstein polaron model for
molecular crystals \cite{holstein}. In an optics context, DNLS describes wave motion in
coupled nonlinear waveguides. In the basic nonlinear coupler model introduced in \cite{jensen}
two waveguides made of similar optical material are embedded in a different host material. DNLS,
and generalizations, have also appeared in studies of nonlinear electrical lattices \cite{marquie}.

As a concrete algebraically simple example, in the rest of this section we will concentrate on the
case of $\delta=4$. Qualitatively similar results are expected for generic $\delta$. The linear
stability analysis is treated for generic $\delta$ in appendix \ref{linear}. For most purposes we follow
closely the analysis of \cite{HT} which we recommend for additional details.

\subsubsection{The general structure of solutions}

The recursion relation \eqref{arraysbc} (with the ansatz $\delta=4$)
\beq
\label{arraysbf}
\tilde g\varphi_n+\varphi_{n-1}+\varphi_{n+1}+\frac{\tilde s}{2}\varphi_n |\varphi_n|^2=0
\eeq
can be viewed as a four-dimensional mapping from $\IC^2 \to \IC^2$. Using polar coordinates
\beq
\label{arraysbg}
\varphi_n =r_n e^{i \theta_n}
\eeq
we obtain the following two sets of equations
\beq
\label{arraysbi}
r_{n+1} \cos(\Delta \theta_{n+1})+r_{n-1} \cos (\Delta \theta_n)=
-\left( \tilde g+\frac{\tilde s}{2}r_n^2\right)r_n
~,
\eeq
\beq
\label{arraysbj}
r_{n+1}\sin (\Delta \theta_{n+1})-r_{n-1} \sin(\Delta \theta_{n-1} )=0
~,
\eeq
where
\beq
\label{arraysbk}
\Delta \theta_n \equiv \theta_n-\theta_{n-1}
~.
\eeq
Equation \eqref{arraysbj} is equivalent to the conservation of current
\beq
\label{arraysbl}
J\equiv r_n r_{n-1} \sin(\Delta \theta_n)
~.
\eeq

It is convenient to introduce the real-valued variables
\begin{subequations}
\beq
\label{arraysbma}
x_n \equiv =\varphi_n^* \varphi_{n-1}+\varphi_n \varphi_{n-1}^*=2r_n r_{n-1} \cos (\Delta \theta_n)
~,
\eeq
\beq
\label{arraysbmb}
y_n \equiv i\left ( \varphi_n^* \varphi_{n-1}-\varphi_n \varphi_{n-1}^* \right)=2 J
~,
\eeq
\beq
\label{arraysbmc}
z_n \equiv |\varphi_n|^2-|\varphi_{n-1}|^2=r_n^2-r_{n-1}^2
~.
\eeq
\end{subequations}
In terms of these variables the recursion equations \eqref{arraysbf} become
\begin{subequations}
\beq
\label{arraysbna}
x_{n+1}+x_n=-\left( \tilde g+\frac{\tilde s}{2}(w_n +z_n)\right) (w_n +z_n)
~,
\eeq
\beq
\label{arraysbnb}
z_{n+1}+z_n=\frac{1}{2} \frac{x_{n+1}^2-x_n^2}{w_n+z_n}
~,
\eeq
\beq
\label{arraysbnc}
w_n=\sqrt{x_n^2+z_n^2 +4 J^2}
\eeq
\end{subequations}
thus reducing our 4D map to a 2D map $\MM:\IR^2 \to \IR^2$.

This map depends on two parameters: $(\tilde g,\tilde s)$. The dependence on $J$ can be scaled
away by setting
\beq
\label{arraysbo}
x_n \to 2J x_n~, ~~ z_n \to 2J z_n~, ~~ \tilde s\to 2J \tilde s
\eeq
so that $w_n=\sqrt{1+x_n^2+z_n^2}$.
A linear stability analysis shows (see appendix \ref{linear} for details) that, depending on the
precise parameters, there are both bounded and diverging solutions. In certain regimes, $e.g.$ when
\beq
\label{arraysboa}
\tilde s>0~,~~ \tilde g<2~,  ~~
-\frac{2(\tilde g+2)}{\tilde s}<|\varphi_n|^{\delta-2}<\frac{2(2-\tilde g)}{\tilde s(\delta-1)}
\eeq
the solutions are regular and bounded and the Lyapunov exponent vanishes \cite{HT}.
Recall that the original parameter $s$ that appears in the AdS/CFT context \eqref{condam}
is positive, but $\tilde s=\frac{s}{h}$ can be both positive or negative. Moreover, one can tune $s$
freely by adding on the field theory dual quartic-trace interactions of the form $|\OO|^4$.

The general structure of the space of solutions is organized by a hierarchy of periodic orbits
surrounded by quasi-periodic orbits \cite{HT}. The periodic orbits can be traced on the intersection
of any two symmetry lines $\SS_0^n=\MM^n \SS_0$, $\SS_1^n=\MM^n \SS_1$, $n=0,1,\ldots$
\cite{Lichtenberg,mackay}, where the fundamental symmetry lines are defined as
\begin{subequations}
\beq
\label{arraysbpa}
\SS_0~:~~z=0
~,
\eeq
\beq
\label{arraysbpb}
\SS_1~:~~ x=-\frac{1}{2}\left( \tilde g+\frac{\tilde s}{2} (w+z)\right) (w+z)
~.
\eeq
\end{subequations}
For example, in the intersection of the lines $\SS_0$ and $\SS_1$ one locates the fixed point
$(x=x_*,z=0)$ with
\beq
\label{arraysbq}
x_*=-\frac{1}{2}\left( \tilde g+\frac{\tilde s}{2} \sqrt{1+x_*^2}\right) \sqrt{1+x_*^2}
~.
\eeq

The linear stability of an orbit with period $q$ is conveniently characterized by the value of Greene's
residue \cite{greene}
\beq
\label{arraysbr}
\rho=\frac{1}{4}\left [ 2- {\rm Tr} \left( \prod_{n=1}^q D\MM^{(n)} \right) \right]
\eeq
where $D\MM$ is the linearization of the map $\MM$. The period orbit is linearly stable when
$0<\rho<1$ (elliptic periodic orbit) and unstable when $\rho>1$ (hyperbolic with reflection) or
$\rho<0$ (hyperbolic).

In the case of the fixed point $(x_*,0)$ the residue is
\beq
\label{arraysbs}
\rho=1-\frac{1}{4}\left( \tilde g+\frac{\tilde s}{2} \sqrt{1+x_*^2} \right)(\tilde g+\tilde s \sqrt{1+x_*^2})
~.
\eeq
For $\tilde s=0$ eq.\ \eqref{arraysbq} has one root, $x_*^2=\frac{\tilde g^2}{4-\tilde g^2}$, which is
real when $|\tilde g|<2$.
In that case, $0<\rho=1-\frac{1}{4}\tilde g^2<1$, so one obtains an elliptic fixed point. This conclusion
continues to hold for generic $\tilde s>0$ and $|\tilde g|<2$.

In the regime of elliptic stability the fixed points $(x_*,0)$ form the largest basins of stability among
all elliptic orbits. These stable orbits, which include both periodic and quasi-periodic solutions,
encircle the fixed point forming the main island on the map plane. The quasi-periodic orbits,
which lie on closed curves (the Kolmogorov-Arnold-Moser (KAM) tori), densely fill the island.
One can show that the map $\MM$ is topologically equivalent to an area-preserving map ensuring
the existence of such KAM-tori near the symmetric elliptic fixed points \cite{arnold}.

An illustration of the main island of elliptic orbits around the fixed point $(x_*,0)$ for $\tilde g=1.6$,
$\tilde s=0.1$ can be found in plot $(a)$ of Fig.\ \ref{complex}. Outside this island one finds
regular quasiperiodic orbits and orbits that diverge.

\FIGURE[t]{
\vspace{.4cm}
\centerline{\includegraphics[width=7cm,height=4cm]{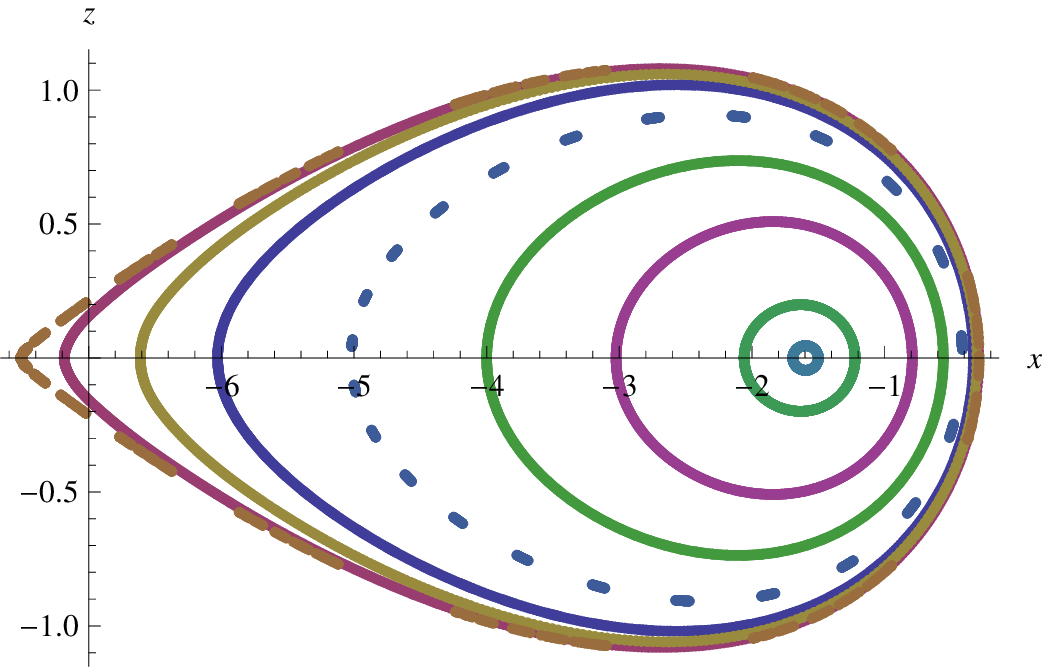} ($a$) \hspace{0.5cm}
\includegraphics[width=7cm,height=4cm]{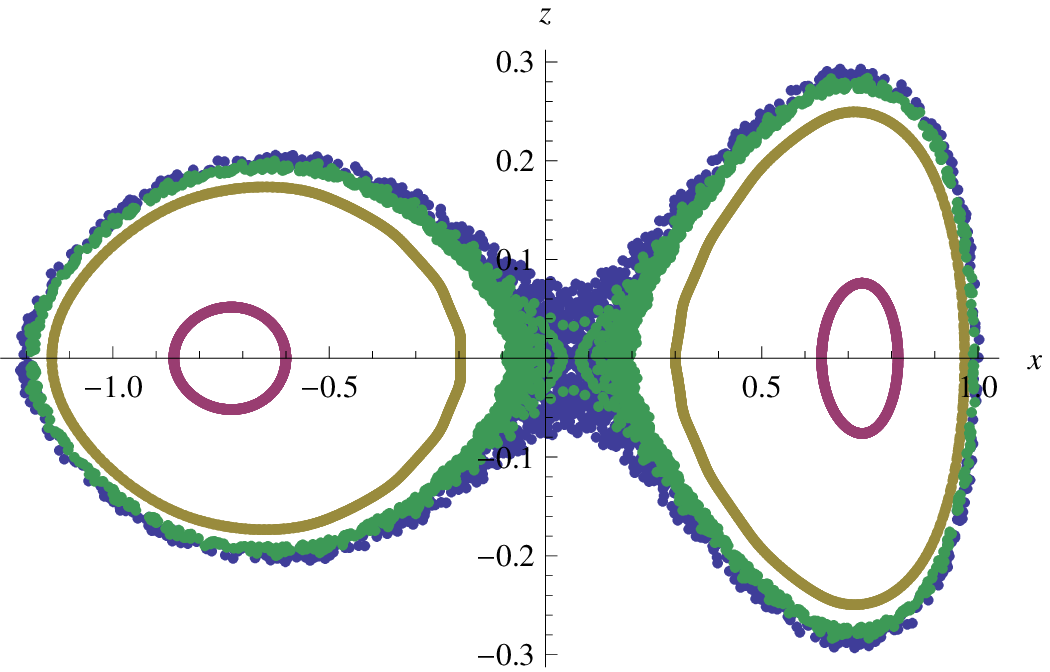} ($b$)}
\caption{\small \it Plot (a) depicts in the $(x,z)$-plane the main island of elliptic orbits that develops
around the elliptic fixed point $(x_*,0)$ with $x_*\simeq -1.59355$ for $\tilde g=1.6$, $\tilde s=0.1$.
Plot (b) depicts a characteristic example of period-doubling bifurcation for $\tilde g=-0.62$,
$\tilde s=1$. Orbits with different initial conditions are depicted with different colors.}
\label{complex}
}

Interesting changes in the structure of the solution space can occur as we vary the parameters of
the system, here $(\tilde g,\tilde s)$. In particular, the residue of periodic orbits can change. When
the residue changes from a positive value to a negative value then a tangent bifurcation occurs
where an elliptic point converts into a hyperbolic point. Whenever the residue exceeds the value
of one from below a stable elliptic orbit converts into an unstable hyperbolic point with reflection
accompanied by the creation of two new stable elliptic points. This kind of bifurcation is known as
period-doubling bifurcation ---a 1-period orbit (fixed point) converts into a 2-period orbit.
The new elliptic orbits remain stable until another period-doubling bifurcation occurs. After a cascade
of such bifurcations local chaos appears.

In our system, the residue $\rho$ in eq.\ \eqref{arraysbs}, is always less than one assuming $\tilde s>0$,
$\tilde g>0$. In that case, only tangent bifurcations can occur and global chaos can arise through
the so-called resonance overlap. Period-doubling bifurcation can instead occur when $\tilde g<0$.
Specifically, in the range $\tilde g<-\frac{\tilde s}{2}<0$ a new period-2 orbit is generated at the points
\beq
\label{arraysbt}
x_\pm=\pm \sqrt{4\frac{\tilde g^2}{\tilde s^2}-1}~, ~~z=0
~.
\eeq
The map $\MM$ acts on these points by sending $(x_\pm,0)\to (x_\mp, 0)$. The residue of
the new solution is positive
\beq
\label{arraysbu}
\rho=\frac{1}{2}\left(\tilde g^2-\frac{\tilde s^2}{4}\right)
~.
\eeq
As we decrease $\tilde g$ further the period-2 orbit loses its stability and a new period-doubling
bifurcation occurs which gives rise to a period-4 orbit. This cascade terminates at a
critical parameter (see \cite{HT} and references therein)
\beq
\label{arraysbv}
\tilde g_\infty=-\left(\tilde s+\sqrt{\frac{\tilde s^2}{4}-|C_\infty|}\right)~, ~~ C_\infty \simeq -1.2656
\eeq
which is called the accumulation point. At that point local chaos appears.

A characteristic example of period-doubling bifurcation is depicted in plot $(b)$ of Fig.\ \ref{complex}.
At $\tilde g=-0.62$, $\tilde s=1$, the originally stable fixed point of plot $(a)$ has become unstable
and two new elliptic fixed points have been created giving rise to elliptic period-2 orbits.

Besides the issue of linear stability, that was discussed above, one can also ask about the
local and global thermodynamic stability of the above solutions. Local thermodynamic stability
requires a positive definite Hessian of the multi-gravity energy functional \eqref{condaq}
(see also \eqref{condan}). It would be interesting to examine the extent to which local thermodynamic
stability is equivalent to linear stability. On the other hand, global thermodynamic stability implies that
the more stable solutions have less energy. In the following subsections \ref{periodic} and \ref{finite}
we will see that chains with a finite number of sites have a discrete set of solutions. In that case,
the solutions with the minimum energy are thermodynamically favored. We hope to return to a more
detailed examination of these issues in the future.

\subsubsection{Periodic boundary conditions}
\label{periodic}

From the above discussion it should be clear that, for given values of the parameters and period,
one is left with at most a discrete finite set of solutions to the recursion relations \eqref{arraysbf} in the
case of periodic boundary conditions. Indeed, in regimes that allow for elliptic periodic
orbits the choice of a prescribed period picks the sequence of $\varphi_n$'s in general uniquely.
In other regimes of parameters periodic solutions do not even exist. Moreover,
through period-doubling bifurcation it is interesting to note that it is possible to have periodic
solutions where the $\varphi_n$'s arrange themselves in more than one different domains of
values.

\subsection{One or two boundaries}
\label{finite}

If there is a boundary, say at $n=0$ with $n$ valued only on non-negative integers,
the $n=0$ version of the equation \eqref{arraysbf} is modified to
\beq
\label{finiteaa}
\tilde g\varphi_0+\varphi_1 +\frac{\tilde s}{2}\varphi_0 |\varphi_0|^{\delta-2}=0
~.
\eeq
In that case, the whole solution is fixed by the choice of one parameter, for example $\varphi_0$.
If the chain has finite size and there is also a second boundary, then the analog of \eqref{finiteaa}
at the second boundary will fix $\varphi_0$ as well and the solution will be discretely unique
and expressed completely in terms of the parameters of the system $g$, $s$, $h$. A
specific example of this situation appeared in the dimer case of section \ref{dimer}.

A special consequence of eq.\ \eqref{finiteaa} is the fact that $\varphi_n$ are all real-valued
up to a common $n$-independent phase. Equivalently, the phases $\theta_n$ in \eqref{arraysbg} are
all equal modulo $\pi$. This property can be deduced by using the boundary equation
\eqref{finiteaa} to compute the conserved ($i.e.$ $n$-independent) current $J$ \eqref{arraysbmb}.
One finds $J=0$ from which the above statement follows immediately.

Taking care to satisfy the conditions from the boundary equations one can proceed as before
to analyze the solutions. Depending on the precise parameters and the boundary value $\varphi_0$
one finds again in the case of a semi-infinite chain regular or diverging solutions. The regular
solutions can be periodic or quasi-periodic. In the case of a chain with finite size only a discrete
subset of the above regular solutions survives. We are not aware of any tractable analytic method
that determines these regular solutions for generic sets of parameters.

\subsubsection{Solitons and pinning of superconductivity}

It is known that the DNLS equation admits also another interesting kind of solutions: soliton and kink
solutions. A detailed discussion of these solutions in the real domain and related references can be
found in \cite{HT}  whose main points can be summarized briefly as follows. A priori one might expect
that the DNLS equation does not admit such solutions. Soliton-like solutions are typically associated
to integrable systems and DNLS is not integrable. It exhibits irregular chaotic behavior which in
principle may prevent perfect localization. Nevertheless, it can be shown that non-integrability and
discreteness appropriately combine to make such solutions possible.

The solutions of interest have the following characteristics. They are solutions where the amplitude
$\varphi_n$ is exponentially localized around a single site, say at $n=0$. Following the nomenclature
of \cite{HT} one can distinguish between two situations:
\begin{itemize}
\item[$(1)$] {\bf Bright solitons}: in this case, $|\varphi_n|>|\varphi_{n+1}|$ for $n>0$ and
$|\varphi_n|<|\varphi_{n+1}|$ for $n<0$ with $\lim_{|n|\to \infty}|\varphi_n|=0$.
\item[$(2)$] {\bf Dark solitons}: in this case $|\varphi_n|<|\varphi_{n+1}|$ for $n>0$
and $|\varphi_{n+1}|>|\varphi_n|$ for $n<0$ with $\lim_{|n|\to \infty}|\varphi_n|>0$. It turns out
that $\lim_{n\to +\infty}\varphi_n=-\lim_{n\to -\infty}\varphi_n$, so these solutions are really kink
solutions.
\end{itemize}

In our context, where each site labeled by an index $n$, models a (1+1)- or (2+1)-dimensional
layer of a superconducting material, such configurations would correspond in case (1) to
situations where in a chain of layers the interlayer interactions work in such a way that energy
and superconductivity are strongly localized around a central site. In case (2) the opposite happens.
Energy and superconductivity are modulated in such a way that they are almost uniform along the
chain except around a central site where the condensate vanishes as it changes sign and
superconductivity becomes very weak.

In a continuum limit (see next subsection) the above configurations appear to reconstruct a junction
of three materials with one dimension higher. In case (2) we recover a configuration that is very similar
to the dark soliton of \cite{Keranen:2009ss} and reminds of an SNS junction of (2+1)- or (3+1)-
dimensional superconductors. From this point of view the configuration of case (1) resembles a
junction of two materials in the normal state separated by a thin superconducting layer in the middle.

Since $\varphi_n$ are now real it is convenient to view the equations \eqref{arraysbf} as a
two-dimensional map $\widetilde \MM~:~\IR^2 \to \IR^2$ by defining a new set of $\IR^2$
coordinates $(x_n,y_n)=(\varphi_n,\varphi_{n-1})$. Then,
\beq
\label{finiteab}
\widetilde \MM~:~ \bigg \{
\begin{array}{c}
x_{n+1}=-\left( \tilde g+\frac{\tilde s}{2} x_n^2 \right) x_n-y_n \\
y_{n+1}=x_n
\end{array}
~.
\eeq
The identification of the soliton-like solutions is closely related to the structure of the fixed points
of this map.

The fixed points, which by definition obey the relation $x=y$, are located at
\beq
\label{finiteac}
x_0=0~, ~~ x_\pm =\pm \sqrt{-\frac{2(\tilde g+2)}{\tilde s} }
~.
\eeq
The $x_\pm$ fixed points exist only when ${\rm sgn}(\tilde g+2)=-{\rm sgn} (\tilde s)$.
Greene's residue $\rho$ for the fixed point at the origin is \cite{HT}
\beq
\label{finitead}
\rho=\frac{1}{4}(\tilde g+2)
~.
\eeq
Consequently, for $|\tilde g|<2$, we obtain $0<\rho<1$ which implies that the origin is a stable elliptic
fixed point encircled by stable elliptic orbits. When in addition $\tilde s<0$ the fixed points $x_\pm$
are unstable hyperbolic fixed points.

For $|\tilde g|>2$ we may distinguish between the following two cases:
\begin{itemize}
\item[$(i)$] $\tilde g<-2$, $\tilde s>0$. In that case $\rho<0$ and the origin becomes an
unstable hyperbolic point. The points $x_\pm$ are stable elliptic fixed points.
\item[$(ii)$] $\tilde g>2$, $\tilde s<0$. The fixed point at the origin becomes unstable and through
period-doubling bifurcation a new period-2 orbit appears located on the line $x=-y$.
\end{itemize}

Before proceeding to explain the main idea underlying the existence of soliton-like solutions
it will be useful to introduce some language which is common in the study of dynamical systems.

A set $\WW$ is called an invariant manifold of a dynamical system if for any point $x\in \WW$
the dynamical evolution of $x$ for any amount of time $t$ continues to belong in $\WW$.
Every fixed point $p$
comes with its invariant manifolds. Such manifolds are called stable, and denoted as $\WW^s(p)$,
if all points that belong on them approach asymptotically the fixed point $p$ under dynamical evolution
(namely $p$ is an attractor on $\WW^s(p)$). In contrast, an invariant manifold is called unstable, and
denoted as $\WW^u(p)$, if all points that belong to it move asymptotically away from the fixed point
$p$ under dynamical evolution (in other words, $p$ is a repellor on $\WW^u(p)$).

For generic non-integrable maps it is known that the stable and unstable invariant manifolds of
hyperbolic fixed points cross each other. Points that reside on the intersection of stable and
unstable invariant manifolds of the same fixed point are called {\it homoclinic} points.
Accordingly, points that reside on the intersection of stable and unstable invariant manifolds of
two different hyperbolic fixed points are called {\it heteroclinic} points.

Having made this short introduction, we are now in position to describe what happens in our specific
system provided by the map \eqref{finiteab}. First consider the case $(i)$ with $\tilde g<-2$, $\tilde s>0$.
The origin is an unstable hyperbolic point. Moving along an orbit on an unstable manifold $\WW^u$ of
the origin and then crossing through a homoclinic point to a stable manifold $\WW^s$ gives rise to a
soliton-like solution of the type (1) above. An explicit computation of the stable and unstable manifolds
in the case of DNLS as well as specific examples can be found in \cite{HT} (see, for instance,
Fig.\ 7 in \cite{HT}).

In case $(ii)$ with $\tilde g>2$, $\tilde s<0$ one obtains a similar soliton-like solution, but with the
new feature that adjacent amplitudes $\varphi_n$, $\varphi_{n+1}$ have alternating signs. Such
solutions are known as staggered solitons \cite{cai}. The solitons of the previous paragraph are also
known as unstaggered solitons.

Finally, in the case of $|\tilde g|<2$, $\tilde s<0$ one can consider heteroclinic orbits connecting the
two unstable hyperbolic fixed points $x_\pm$. These are kink solutions of the type (2) above.

\subsection{Continuum limit and the Gross-Pitaevskii equation}
\label{continuum}

It is interesting to consider the continuum limit of the chain configuration described above. In this limit
\beq
\label{arraysea}
M\to \infty~, ~~\frac{n}{M} \to x~, ~~ \varphi_n \to \varphi(x)~, ~~
g \to \frac{G}{M^2}~, ~~ s\to \frac{S}{M^2}
\eeq
with the new parameters $x$, $G$, and $S$ kept finite.
In this limit the recursion relations \eqref{arraysbc} turn into the second order non-linear
Schr\"odinger differential equation
\beq
\label{arrayseb}
\varphi ''+\varphi\left( G+\frac{S}{2}|\varphi|^{\delta-2} \right)=0
\eeq
where $' \equiv \frac{d}{dx}$.

For $\delta=4$ we recover directly a well-known equation in the context of superfluids; the
Gross-Pitaevskii equation \cite{Gross,pita} which gives a coarse-grained description
of superfluids at long wavelengths. The GP equation is typically relevant for weakly-interacting
Bose-Einstein condensates or strongly bound fermionic superfluids at low temperature. It is
interesting that in our formalism, which has a radically different point of departure, the same
description emerges naturally (as a candidate description of strongly coupled superconductor
physics) out of a framework designed specifically to deal with layered structures.
It would be worth exploring further parallels that may exist between our formalism (and the
generalizations of the GP equation that it suggests) and the known applications of the GP equation to
superfluidity and superconductivity.

For simplicity and concreteness let us continue to concentrate on the case of $\delta=4$. Two
well-known solutions of this equation are:
\begin{itemize}
\item[(1)] {\bf Bright solitons}: for $G<0, S>0$
\beq
\label{brightaa}
\varphi(x)=\pm \sqrt{-\frac{4G}{S}} \frac{1}{\cosh\left(\sqrt{-G}\, x\right)}
~,
\eeq
\item[(2)] {\bf Dark solitons}: for $G>0, S<0$
\beq
\label{darkaa}
\varphi(x)=\pm \sqrt{\frac{2G}{S}} \tanh \left( \sqrt{\frac{G}{2}}x\right)
~.
\eeq
\end{itemize}
For real $\varphi$ we can find a more general class of solutions expressed in terms of the
Jacobi elliptic function $sn(u|m)$
\beq
\label{arraysecJ}
\varphi(x)=\pm i \sqrt{\frac{2(G-\sqrt{G^2+S C})}{S}} ~
sn\left((x+x_0) \sqrt{\frac{G+\sqrt{G^2+SC}}{2}} \, \Bigg | \,
\frac{G-\sqrt{G^2+S C}}{G+\sqrt{G^2+S C}}\right)
~.
\eeq
$C$ and $x_0$ are integration constants.

Reinstating the phase \eqref{arraysbb} we obtain
\beq
\label{arraysecP}
\alpha(x)=e^{i x \theta}\varphi(x)
\eeq
where $\theta$ is the finite rescaled version of the angular inter-layer coupling $\vartheta$
\beq
\label{arraysed}
\theta \equiv M \vartheta
~.
\eeq

At a boundary point $x_b$ the discrete equation \eqref{finiteaa} becomes in the continuum limit
\eqref{arraysea} a Dirichlet boundary condition
\beq
\label{arraysee}
\varphi(x_b)=0
~.
\eeq

By varying the parameters of the solution \eqref{arraysecJ} one obtains qualitatively different
behaviors. For concreteness, set $C=1$, $x_0=0$. For $G>0$ and generic $S\neq 0$ one finds
periodic sine-like solutions like the one depicted in plot $(a)$ in Fig.\ \ref{sols}. For $G<0$ and $S>0$
the generic periodic solution, that appears as plot $(b)$ in Fig.\ \ref{sols}, can be suitably
tuned to obtain the bright-soliton solution \eqref{brightaa} (see also plot $(c)$ in
Fig.\ \ref{sols}). For $G>0$ and $S<0$ one can find kink configurations by choosing the
parameters $G,S$ so that the second argument in the Jacobi sine function becomes 1.
In that case, we recover the dark soliton solution \eqref{darkaa} (an example of such a
configuration appears in plot $(d)$ of Fig.\ \ref{sols}).

\FIGURE[t]{
\vspace{.4cm}
\centerline{\includegraphics[width=7cm,height=4cm]{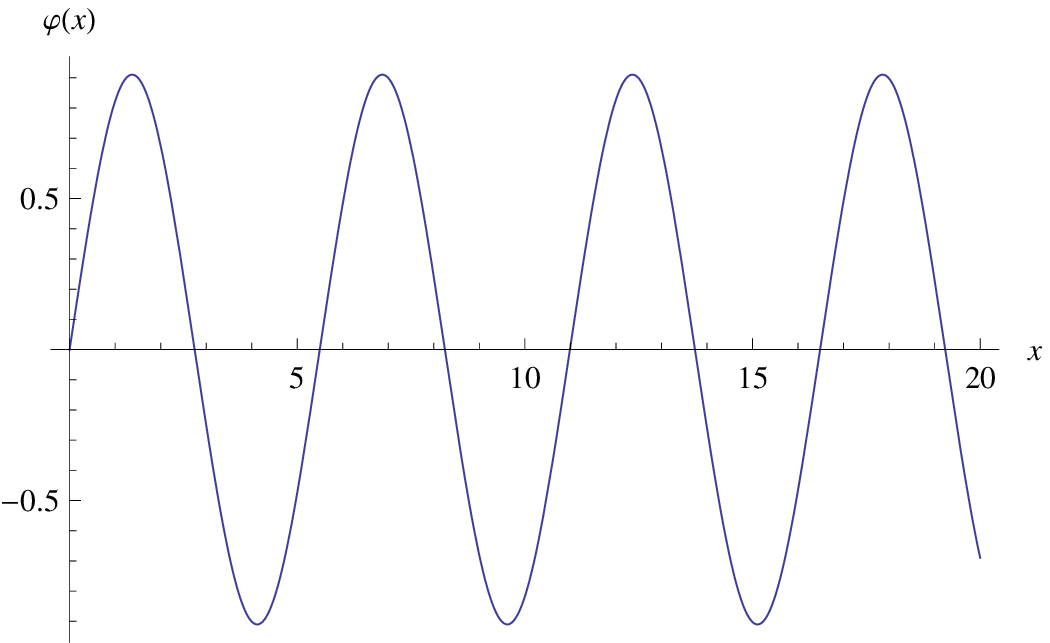} ($a$) \hspace{0.5cm}
\includegraphics[width=7cm,height=4cm]{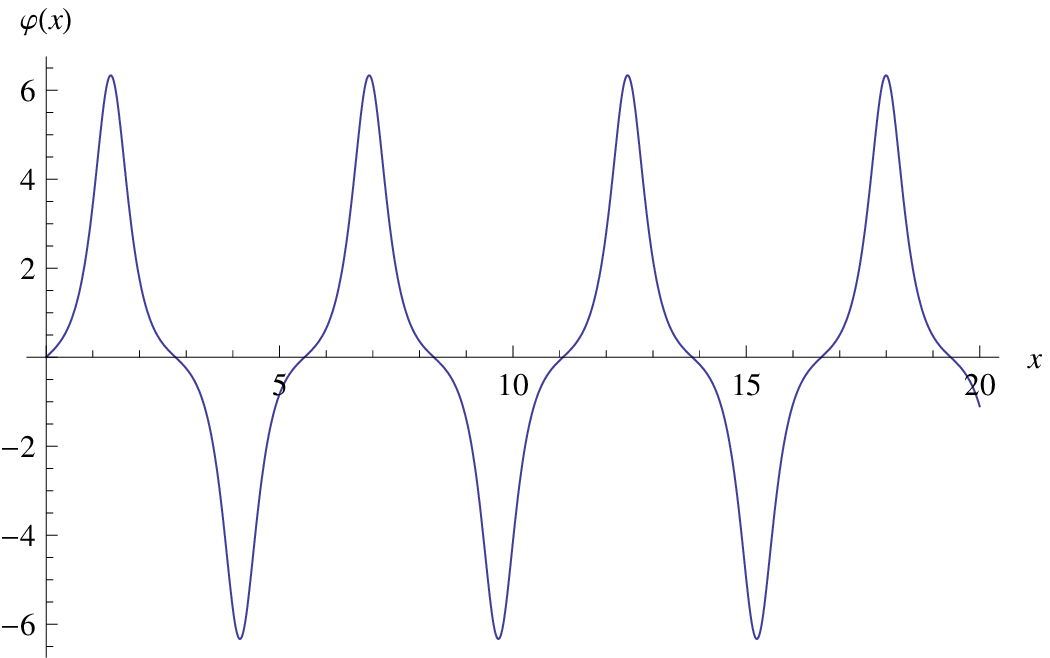} ($b$)}
\centerline{\includegraphics[width=7cm,height=4cm]{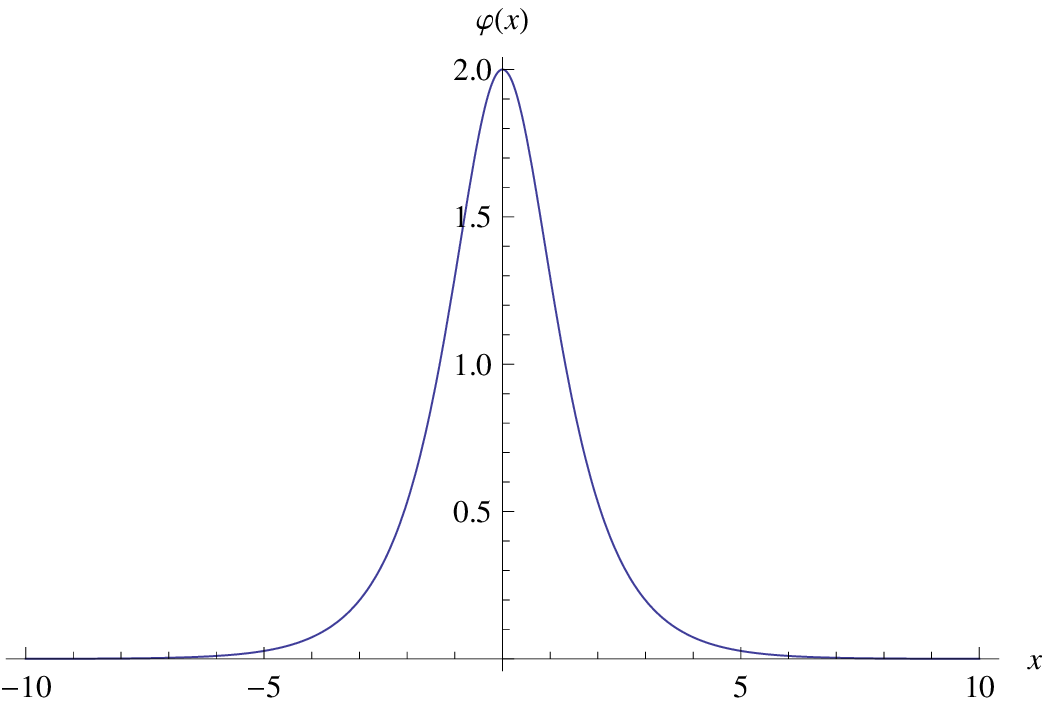} ($c$) \hspace{0.5cm}
\includegraphics[width=7cm,height=4cm]{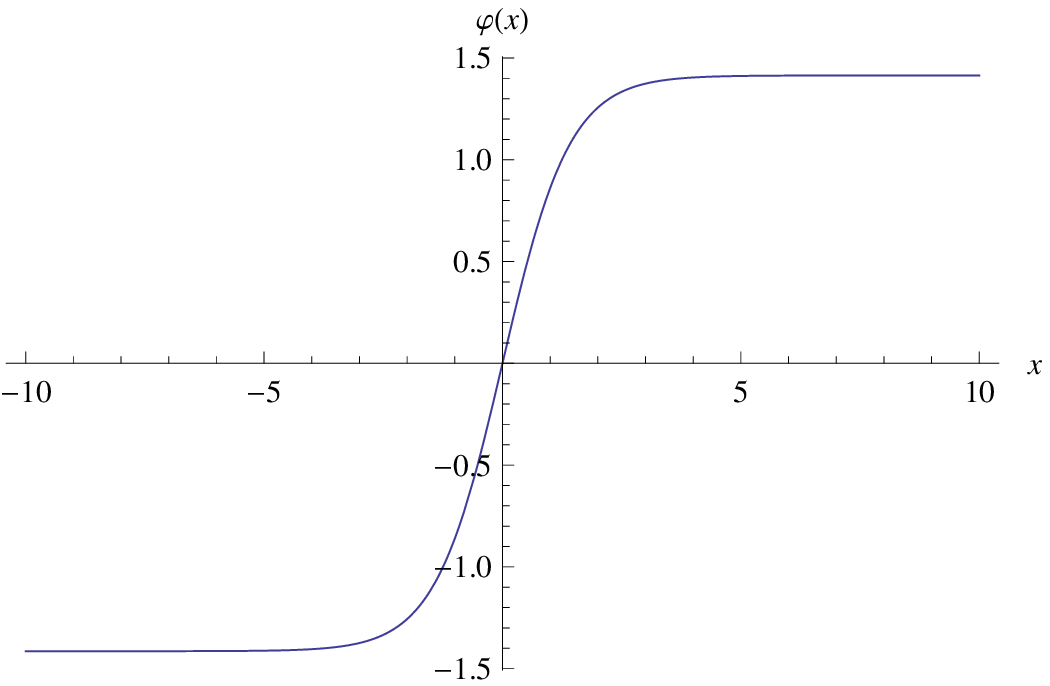} ($d$)}
\caption{\small \it Plot (a) depicts a generic sine-like periodic configuration for $G>0$. The specific
plot has $G=S=1$. Plot (b) depicts a generic periodic configutation for $G<0$, $S>0$ (in this
particular case $G=-10, S=1$). By suitably tuning $G,S$ one obtain the bright-soliton of eq.\
\eqref{brightaa}, here depicted in plot (c) for $G=-1, S=1$.
A dark soliton (or kink) solution can be obtained by tuning $G,S$ so that the second
argument in the Jacobi sine function becomes 1. An example of this case appears in plot (d)
for $G=-S=1$.}
\label{sols}
}

By suitably truncating any of the solutions depicted in plots $(a), (b)$, or $(c)$ of Fig.\ \ref{sols} within
an interval bounded by the location of two zeros of $\varphi$ one obtains trivially a finite-size Josephson
junction chain in a continuum limit. For the plot $(d)$ there is a single point where $\varphi$ vanishes
(the core of the kink solution), hence solutions of this type that are consistent with the Dirichlet
boundary conditions \eqref{arraysee} do not exist.

\subsection{Josephson current across a chain}
\label{chaincurrent}

Following the discussion of subsection \ref{Jcurrent} we can write the total current across the 
chain (evaluated across two adjacent superconductors at positions $n-1$, $n$) as a sum of two 
contributions 
\beq
\label{chaincurrentaaa}
J_{tot}=\langle J_{n-1,n} \rangle + J^{josephson}_{n-1,n}
~.
\eeq
$J_{n-1,n}$ is the current operator 
\beq
\label{chaincurrentaab}
J_{n-1,n}=i(\OO_{n-1} \OO_n^\dagger-\OO_{n-1}^\dagger \OO_n)
\eeq
associated to the charge-transferring part of the $(n-1,n)$ interlayer interaction in 
eq.\ \eqref{arraysaa}, and $J^{josephson}_{n-1,n}$ 
is the Josephson current associated to the backreaction of the system discussed in subsection 
\ref{Jcurrent}. The total current $J_{tot}$ is the total conserved current of the system, which, 
up to a potential multiplicative constant that we will keep implicit, equals the quantity $J$ 
in eq.\ \eqref{arraysbl}. This is a site-independent quantity. In a chain with finite length the boundary
conditions set $J=0$ and $\langle J_{n-1,n}\rangle = - J^{josephson}_{n-1,n}$ as in the two-site
system of subsection \ref{Jcurrent}. More generally, however,
\beq
\label{chaincurrentaac}
J=\langle J_{n-1,n}\rangle +J^{josephson}_{n-1,1}
~.
\eeq

Using the definition \eqref{chaincurrentaab} we find that the current
\beq
\label{chaincurraa}
\langle J_{n-1,n}\rangle 
=i(\alpha_{n-1}^* \alpha_n-\alpha_{n-1}\alpha_n^*)=i(e^{i\vartheta} \varphi_{n-1}^* \varphi_n
-e^{-i\vartheta} \varphi_{n-1} \varphi_n^*)
~.
\eeq
In terms of the polar coordinates \eqref{arraysbg} we further obtain
\beq
\label{chaincurrab}
\langle J_{n-1,n} \rangle =
-2r_{n-1}r_n \sin(\vartheta+\Delta \theta_n)
=-2J \frac{\sin(\vartheta+\Delta \theta_n)}{\sin(\Delta \theta_n)}
\eeq
where $\Delta \theta_n$ and $J$ were defined in eqs.\ \eqref{arraysbk} and \eqref{arraysbl}
respectively.

For a general solution of the recursion equations \eqref{arraysba} and $\vartheta\neq 0,\pi$
this is a link-dependent current. It is periodically or chaotically modulated in periodic or
quasi-periodic solutions. In real solutions, where $\Delta \theta_n=0 \mod \pi$,
\beq
\label{chaincurrac}
\langle J_{n-1,n}\rangle =\pm 2 r_{n-1}r_n \sin\vartheta
~.
\eeq
For instance, in soliton-like solutions this current is very weak except around a central site.

In the special case where $\vartheta=0,\pi$ ($i.e.$ when no external interlayer gauge field is applied), 
eq.\ \eqref{chaincurrab} gives the site-independent current
\beq
\label{chaincurrad}
\langle J_{n-1,n}\rangle =\pm 2J
~.
\eeq
This current vanishes for a chain with boundaries, but can be non-zero in chains without boundaries,
$e.g.$ in a circular chain with periodic boundary conditions. This is a simple example of how the
topology of the network can affect the qualitative features of the configuration.

\subsection{Towards a typical Josephson junction}

\FIGURE[t]{
\vspace{.4cm}
\centerline{\includegraphics[width=12cm,height=7cm]{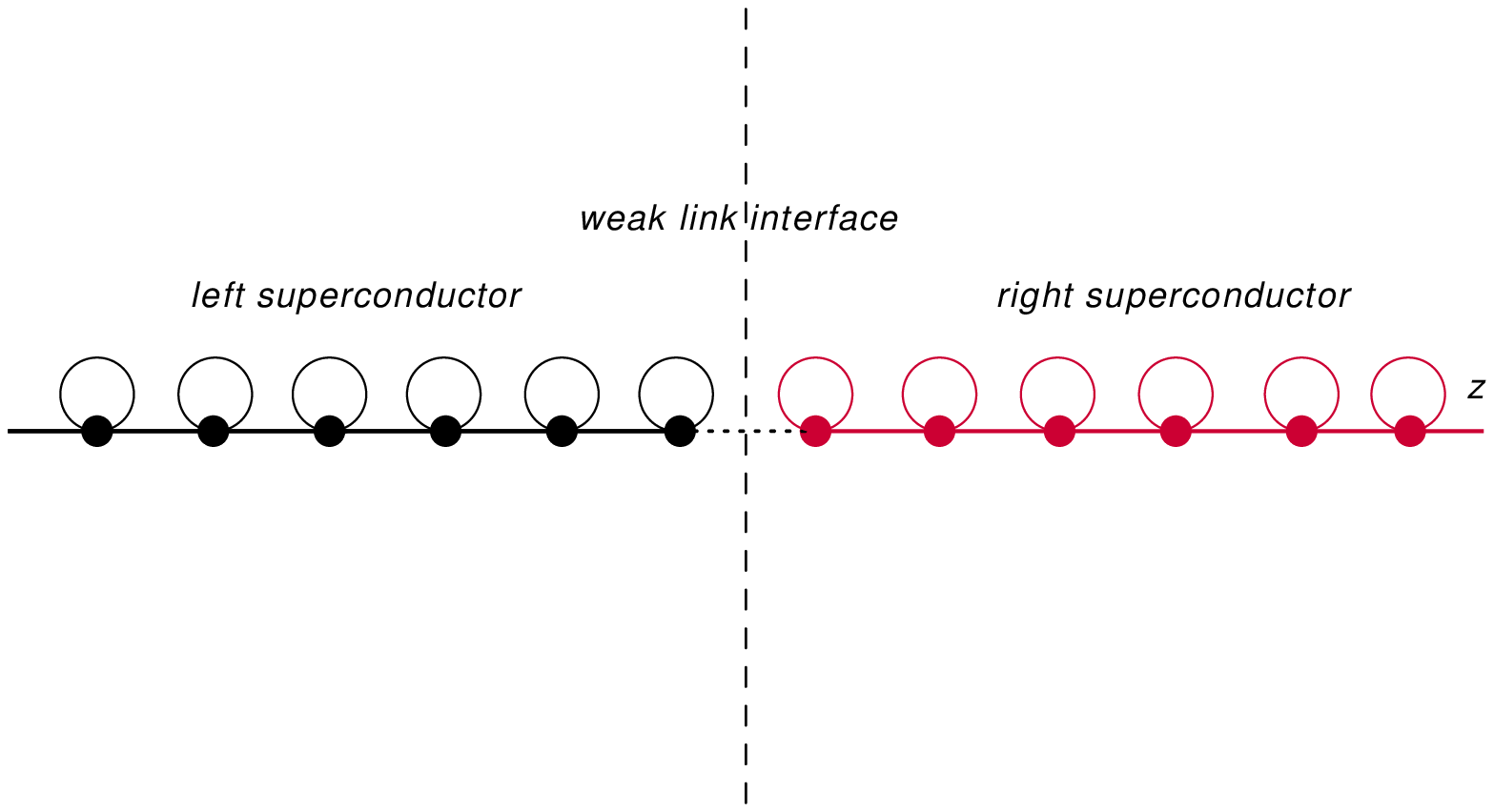}}
\caption{\small \it An (un)conventional JJ constructed as two semi-infinite AdS/CFT arrays
of (2+1)-dimensional holographic superconductors linked to each other at a two-dimensional
weak-link interface.}
\label{networks7}
}

In subsection \ref{Jcurrent} we discussed the similarities and differences between a two-site
system and typical Josephson junctions. Here we discuss how the two-site system can be
extended to look more like the typical Josephson junction. 

Assume we want to describe a junction of two superconductors in three spatial dimensions linked 
weakly across the third direction $z$ at a two-dimensional interface. In previous subsections 
we described how to deconstruct $(3+1)$-dimensional layered superconductors from an array of  
$(2+1)$-dimensional holographic superconductors using linear AdS/CFT arrays. To construct an 
(un)conventional Josephson junction of two superconductors of this type a possible strategy is
described in Fig.\ \ref{networks7}. A layered superconductor on the left (right) is deconstructed as 
an array of cites linked through interactions of the form
\beq
\label{typicalaa}
\sum_n h_{L(R)} \left( \OO_n^{L(R)} \OO_{n+1}^{L(R)\dagger}+ \OO_{n}^{L(R)\dagger} 
\OO_{n+1}^{L(R)}\right)
~.
\eeq
With a real coupling $h_{L(R)}$ no external transverse gauge field is applied along the $z$ direction.
Across the two-dimensional interface the right-most black site of the left chain can be linked to 
the left-most red site of the right chain through a link of a different type depending on the specific
nature of the left and right sites. For $s$-wave holographic superconductors both on the left and 
the right a simple example of a double-trace weak link is 
\beq
\label{typicalab}
h_{link}\left( \OO^L \OO^{R\dagger}+\OO^{L\dagger}\OO^R\right)
~.
\eeq
Then one can solve the analog of the equations \eqref{arraysba} and determine the Josephson 
current as was described in the previous subsection. SNS-type solutions of a uniform array 
with $h_L=h_R=h_{link}$ were described in subsections \ref{infinite}, \ref{finite}, \ref{continuum}
(in a discrete or continuum limit). In general, the asymptotic difference of the phase of the condensates, 
$\Delta \vartheta=\vartheta_L-\vartheta_R$, is a dialed quantity in these systems. 
For conventional SNS or SIS-type JJs we anticipate the presence of a Josephson current that 
follows the sine law relation $I_{\rm max} \sin \Delta\vartheta$. We hope to return to a detailed
survey of such systems in future work.

\section{Outline of future directions}
\label{discussion}

We have proposed a novel holographic way to model the physics of Josephson junctions using
networks of (super)gravity theories on asymptotically AdS spacetimes coupled via mixed boundary
conditions. One of the advantages of this approach, compared to previous holographic
approaches, is the versatility by which it can incorporate many different types of Josephson junctions 
and networks with limitless possibilities in their architecture. For conventional SNS or SIS-type 
superconductors we presented a simple two-site model that exhibits some of the standard 
features of Josephson junction, $e.g.$ the sine relation between the Josephson current and 
condensate phase difference. We explained in what sense this system is different from the typical
Josephson junctions and how one can use AdS/CFT arrays to describe the more typical systems.
We have also seen how a simple network on a chain produces complex dynamics with a variety of 
interesting features.

Our preliminary analysis opens the possibility for a diverse set of calculations and extensions. Some
of the most prominent ones are the following.

\subsection*{\it (a) Finite temperature and a more complete analysis of phenomenological implications}

We have so far considered simple examples of holographic JJs at vanishing temperature and charge
densities. It is of obvious interest to extend the setup to finite temperature using hairy black holes in
designer multi-gravity (see $e.g.$ \cite{Hertog:2005hu} for related work) and to explore possible phase
transitions as we vary the temperature and/or charge densities.

Extending the list of examples it is desirable to consider the explicit properties of other holographic
JJs (or JJNs) built from different types of holographic superconductors ($e.g.$ $s$-wave, or $p$-wave).
For example, it will be interesting to define and study interlayer transport coefficients in such models.
The ultimate goal is to explore the extent to which these constructions reproduce known
phenomenological features of JJNs or layered superconductor physics. For example, it would
be interesting to reproduce previously observed non-sinusoidal current-phase relations in
unconventional JJs.

\subsection*{\it (b) Other network architectures and complex behavior}

It has been pointed by many authors (see $e.g.$ \cite{sodano}) that the architecture of a JJN can have
important implications on the physical properties of the system. It is interesting to explore other
configurations and examine how they affect the collective and local properties of the sites.
Networks with double-trace or higher multi-trace interactions can be constructed. An example
that has been studied previously in the condensed matter literature is the Y-Josephson junction
(see for instance \cite{Ysodano}). The architecture of a Y-Josephson junction network appears in Fig.\
\ref{networks6}.

\FIGURE[t]{
\vspace{.4cm}
\centerline{\includegraphics[width=8cm,height=5.5cm]{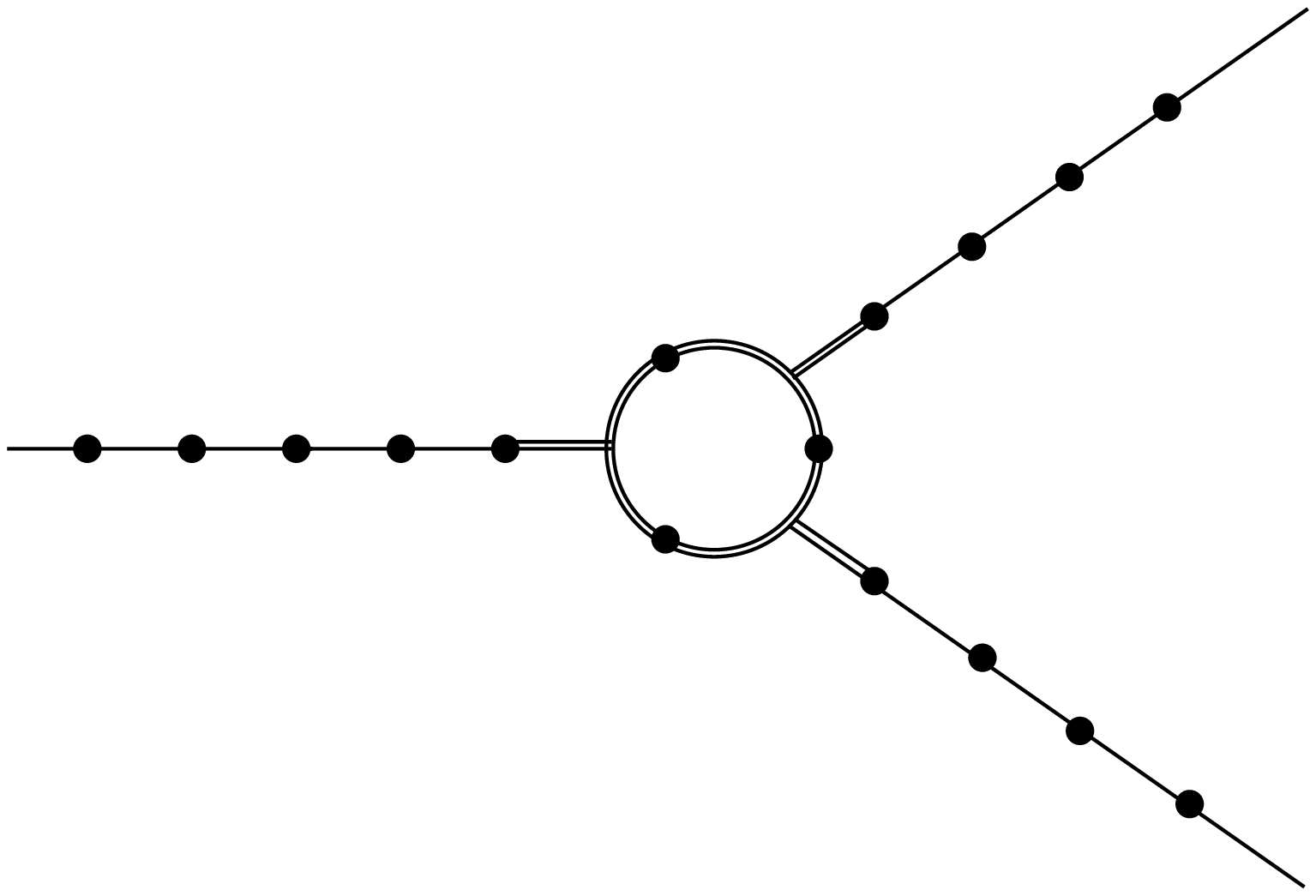}}
\caption{\small \it The architecture of the Y-Josephson junction network.}
\label{networks6}
}

It may also be interesting to explore the existence of vortex solutions in two- or three-dimensional
AdS/CFT lattices. This becomes even more interesting in view of the observed connection to the
GP equation in the continuum limit. A recent discussion of vortex solutions to the GP equation
(and a related AdS/CFT application from a different point of view) can be found in \cite{Keranen:2009re}.

\subsection*{\it (c) Continuous limits and deconstruction}

In subsection \ref{continuum} we considered a continuum limit of a one-dimensional holographic
JJN. In this limit the number of sites is scaled to infinity with an appropriate scaling of the other
parameters of the system to zero. For a specific set of parameters we recover in this limit the
Gross-Pitaevskii equation. It would be interesting to explore further relations between the
generalizations of this equation suggested by our formalism and known applications of
the GP methodology in superfluidity and superconductivity.

In addition, it would be interesting to explore similar continuous limits of other JJNs
with more complicated topology and different ingredients. Such limits may be used to simplify
some aspects of the analysis of the network or in order to attempt a novel deconstruction of one or
more extra spacetime dimensions (more comments on this aspect can be found in
\cite{Kiritsis:2008at}).

\section*{Acknowledgements}\label{ACKNOWL}
\addcontentsline{toc}{section}{Acknowledgements}

We would like to thank Nikos Flytzanis, Christos Panagopoulos and George Tsironis for
explaining pertinent aspects of their work and for providing a useful guide through the vast
condensed matter literature on the subject. VN would also like to thank the Galileo Galilei 
Institute for Theoretical Physics for the hospitality and the INFN for partial support during 
the completion of this work. In addition, this work was partially supported by the
European Union grants FP7-REGPOT-2008-1-CreteHEPCosmo-228644 and
PERG07-GA-2010-268246.

\renewcommand{\theequation}{\thesection.\arabic{equation}}
\addcontentsline{toc}{section}{Appendices}

\section*{Appendices}

\begin{appendix}

\section{On the linear stability of DNLS}
\label{linear}

In this appendix we discuss in more detail the linear stability analysis of the DNLS equation
\eqref{arraysbc}
\beq
\label{linearaa}
\tilde g\varphi_n+\varphi_{n-1}+\varphi_{n+1}+\frac{\tilde s}{2}\varphi_n |\varphi_n|^{\delta-2}=0
~.
\eeq

Introducing a small perturbation $u_n$ around a solution $\varphi_n^{(0)}$
\beq
\label{linearab}
\varphi_n=\varphi_n^{(0)}+u_n
\eeq
we obtain (at first order) the equation
\beq
\label{linearac}
u_{n+1}+u_{n-1}+\left( \tilde g +\frac{\tilde s\delta}{4} |\varphi_n^{(0)}|^{\delta-2} \right)u_n
+\frac{\tilde s(\delta-2)}{4} \varphi_n^{(0)2} |\varphi_n^{(0)}|^{\delta-4} u_n^*=0
~.
\eeq
Next we decompose $\varphi_n^{(0)}$ and $u_n$ into their real and imaginary parts
($x_n$, $y_n$ here should not be confused with the corresponding variables in eqs.\ \eqref{arraysbma},
\eqref{arraysbmb} in the main text)
\beq
\label{linearad}
\varphi_n^{(0)}=X_n+i Y_n~, ~~ u_n=x_n+iy_n
~.
\eeq
For these variables we obtain the following two coupled sets of equations
\begin{subequations}
\bea
\label{linearae}
&&x_{n+1}+x_{n-1}+\left( \tilde g+\frac{\tilde s \delta}{4} (X_n^2+Y_n^2)^{\frac{\delta-2}{2}} \right)x_n
\nonumber\\
&&+\frac{\tilde s(\delta-2)}{4} \left( X_n^2+Y_n^2\right)^{\frac{\delta-4}{2}}
\left( (X_n^2-Y_n^2)x_n+2X_nY_n y_n \right)=0
~,
\eea
\bea
\label{linearaf}
&&y_{n+1}+y_{n-1}+\left( \tilde g+\frac{\tilde s \delta}{4} (X_n^2+Y_n^2)^{\frac{\delta-2}{2}} \right)y_n
\nonumber\\
&&+\frac{\tilde s(\delta-2)}{4} \left( X_n^2+Y_n^2\right)^{\frac{\delta-4}{2}}
\left( -(X_n^2-Y_n^2)y_n+2X_nY_n x_n \right)=0
~.
\eea
\end{subequations}
Introducing the notation
\beq
\label{linearag}
M_n^x=-\left[ \tilde g+ \frac{\tilde s \delta}{4} \left( X_n^2+Y_n^2 \right)^{\frac{\delta-2}{2}}
+\frac{\tilde s (\delta-2)}{4} \left( X_n^2+Y_n^2 \right)^{\frac{\delta-4}{2}}(X_n^2-Y_n^2) \right]
~,
\eeq
\beq
\label{linearai}
M_n^y=-\left[ \tilde g+ \frac{\tilde s \delta}{4} \left( X_n^2+Y_n^2 \right)^{\frac{\delta-2}{2}}
-\frac{\tilde s (\delta-2)}{4} \left( X_n^2+Y_n^2 \right)^{\frac{\delta-4}{2}}(X_n^2-Y_n^2) \right]
~,
\eeq
\beq
\label{linearaj}
N_n=-\frac{\tilde s (\delta-2)}{2} \left( X_n^2+Y_n^2 \right)^{\frac{\delta-4}{2}} X_n Y_n
\eeq
we can rewrite the set of equations \eqref{linearae},  \eqref{linearaf} as a matrix equation
\beq
\label{linearak}
\left(
\begin{array}{c}
 x_{n+1} \\
 x_n \\
 y_{n+1} \\
 y_n
\end{array}
\right)=
\left(
\begin{array}{cccc}
M_n^x   &   -1   &  N_n   &   0 \\
1            &   0     & 0         &   0 \\
N_n       &   0     &   M_n^y  &   -1 \\
0            &   0   &   1 &   0
\end{array}
\right)
\left(
\begin{array}{c}
 x_{n} \\
 x_{n-1} \\
 y_{n} \\
 y_{n-1}
\end{array}
\right)\equiv
{\mathbb J}_n
\left(
\begin{array}{c}
 x_{n} \\
 x_{n-1} \\
 y_{n} \\
 y_{n-1}
\end{array}
\right)
~.
\eeq

The characteristic polynomial for the eigenvalues $\lambda$ of the matrix ${\mathbb J}_n$ is
\beq
\label{linearal}
\lambda^4- \left( M_n^x +M_n^y \right) (1+\lambda^2) \lambda
+\lambda^2 \left( M_n^x M_n^y +2- N_n^2\right)+1=0
~.
\eeq
By further setting
\beq
\label{linearam}
\mu_n =-\tilde g -\frac{\tilde s \delta}{4} (X_n^2+Y_n^2)^{\frac{\delta-2}{2}}
~, ~~
\nu_n=\frac{\tilde s (\delta-2)}{4} (X_n^2+Y_n^2)^{\frac{\delta-4}{2}}
\eeq
we can recast \eqref{linearal} into the more convenient form
\beq
\label{linearan}
\left( \lambda^2-\mu_n \lambda+1\right)^2=\lambda^2 \nu_n^2 \left(X_n^2+Y_n^2\right)^2
\eeq
which yields four roots labeled by two $\IZ_2$ indices $\varepsilon_1,\varepsilon_2=\pm$
\beq
\label{linearao}
\lambda_{\varepsilon_1,\varepsilon_2}=
\frac{\mu_n+\varepsilon_1 \nu_n (X_n^2+Y_n^2)+\varepsilon_2
\sqrt{\left(\mu_n+\varepsilon_1 \nu_n (X_n^2+Y_n^2)\right)^2-4}}{2}
~.
\eeq

The discriminant
\beq
\label{linearap}
\Delta=\left(\mu_n+\varepsilon_1 \nu_n (X_n^2+Y_n^2)\right)^2-4=
\left( \tilde g +\frac{\tilde s}{4}(\delta-
\varepsilon_1 (\delta-2))(X_n^2+Y_n^2)^{\frac{\delta-2}{2}} \right)^2-4
\eeq
can be either positive or negative. When $\Delta<0$, the eigenvalues
$\lambda_{\varepsilon_1,\varepsilon_2}$ lie on the unit circle and the solution is linearly stable.
Since
\beq
\label{linearaq}
\Delta=\Delta_+ \Delta_-
\eeq
with
\beq
\label{linearar}
\Delta_\pm = \tilde g \pm 2
+\frac{\tilde s}{4}(\delta-\varepsilon_1 (\delta-2))(X_n^2+Y_n^2)^{\frac{\delta-2}{2}}
\eeq
$\Delta<0$ requires (by definition $\Delta_+>\Delta_-$)
\beq
\label{linearas}
\Delta_+>0, ~~\Delta_-<0
~.
\eeq

For $\tilde s>0$ the second term on the rhs of eq.\ \eqref{linearar} is always positive
(recall that $\delta>2$ for $\Delta<\frac{d}{2}$). Requiring \eqref{linearas} gives
\beq
\label{linearat}
\tilde g<2~, ~~  -\frac{2(\tilde g+2)}{\tilde s}
<(X_n^2+Y_n^2)^{\frac{\delta-2}{2}}<\frac{2(2-\tilde g)}{\tilde s (\delta-1)}
~.
\eeq
The lower bound on $(X_n^2+Y_n^2)^{\frac{\delta-2}{2}}$ is trivial when in addition
$-2<\tilde g$.

For $\tilde s<0$ the analog of \eqref{linearat} is
\beq
\label{linearau}
-2<\tilde g~, ~~ \frac{2(2-\tilde g)}{\tilde s}< (X_n^2+Y_n^2)^{\frac{\delta-2}{2}}
<-\frac{2(\tilde g+2)}{\tilde s (\delta-1)}
~.
\eeq

\end{appendix}


\end{document}